\documentclass[seceqn,nameyear,dvips]{arxstspdf}
\usepackage{colortbl,multirow,hhline}
\usepackage{flushend}
\usepackage{stfloats}
\usepackage{graphicx}

\volume{25}
\issue{4}
\pubyear{2010}
\firstpage{517}
\lastpage{532}
\doi{10.1214/10-STS344}

\begin{document}
\begin{frontmatter}

\title{Block-Conditional Missing at Random Models for Missing Data}
\runtitle{Block-Conditional MAR Models for Missing Data}

\begin{aug}
\author[a]{\fnms{Yan} \snm{Zhou}\ead[label=e1]{yan.zhou@fda.hhs.gov}},
\author[b]{\fnms{Roderick J. A.} \snm{Little}\corref{}\ead[label=e2]{rlittle@umich.edu}}
\and
\author[c]{\fnms{John D.} \snm{Kalbfleisch}\ead[label=e3]{jdkalbfl@umich.edu}}

\runauthor{Y. Zhou, R. J. A. Little and J. D. Kalbfleisch}

\affiliation{Food and Drug Administration, University of Michigan and University of Michigan}
\address[a]{Yan Zhou is Mathematical Statistician,
Center for Drug Evaluation and Research,
Food and Drug Administration, Silver Spring, Maryland, USA
\printead{e1}.}

\address[b]{Roderick J. A. Little is Professor,
Department of Biostatistics, University of Michigan, Ann Arbor, Michigan, USA
\printead{e2}.}

\address[c]{John D. Kalbfleisch is Professor,
Department of Biostatistics, University of Michigan, Ann Arbor, Michigan, USA
\printead{e3}.}
\end{aug}

\begin{abstract}
Two major ideas in the analysis of missing data are (a) the EM
algorithm [Dempster, Laird and Rubin, \textit{J. Roy. Statist. Soc. Ser. B} \textbf{39}
(\citeyear{r7}) 1--38] for maximum
likelihood (ML) estimation, and (b) the formulation of models for
the joint distribution of the data ${Z}$ and missing data
indicators ${M}$, and associated ``missing at random'' (MAR)
condition under which a model for ${M}$ is unnecessary [Rubin,
\textit{Biometrika} \textbf{63} (\citeyear{r33}) 581--592]. Most
previous work has treated ${Z}$ and ${M}$ as
single blocks, yielding selection or pattern-mixture models
depending on how their joint distribution is factorized. This paper
explores ``block-sequential'' models that interleave subsets of the
variables and their missing data indicators, and then make
parameter restrictions based on assumptions in each block. These
include models that are not MAR. We examine a subclass of
block-sequential models we call block-conditional MAR (BCMAR)
models, and an associated block-monotone reduced likelihood strategy
that typically yields consistent estimates by selectively discarding
some data. Alternatively, full ML estimation can often be achieved
via the EM algorithm. We examine in some detail BCMAR models for
the case of two multinomially distributed categorical variables,
and a two block structure where the first block is categorical and
the second block arises from a (possibly multivariate) exponential
family distribution.
\end{abstract}

\begin{keyword}
\kwd{Block-sequential missing data models}
\kwd{block-conditional MAR models}
\kwd{EM algorithm}
\kwd{categorical data}.
\end{keyword}

\end{frontmatter}

\section{Introduction}\label{1}

Missing values arise in empirical studies for many reasons, including
unavailability of the measurements, respondents refusing to answer
certain items on a questionnaire, and attrition in longitudinal
studies. Complete case (CC) analysis, which omits information in the
cases with missing values, is inefficient and potentially biased,
especially if the subjects included in the analysis are systematically
different from those excluded in terms of one or more key variables.
Approaches that incorporate information in the incomplete cases include
nonresponse weighting (Little and Rubin, \citeyear{r22}, Chapter 3);
multiple imputation (MI), where missing values are replaced by multiple
sets of plausible values (Rubin, \citeyear{r34}; Little and Rubin,
\citeyear{r22}, Chapter 5); weighted estimating equation (WEE) methods
(Lipsitz, Ibrahim and Zhao, \citeyear{r15}); and methods based on the
likelihood for a model for the data, such as maximum likelihood (ML) or
fully Bayes modeling. We focus here on the ML approach, although our
models could also be analyzed using Bayesian or MI methods.

Rubin's (\citeyear{r33}) theory on modeling the missing-data mechanism
was a key development in estimation with incomplete data. Rubin
(\citeyear{r33}) formalized the concept of missing-data mechanisms by
treating the missing-data indicators as random variables and assigning
them a distribution. Specifically, let $Z=(Z_{ij})$ denote a
rectangular $n\times p$ data set; the $i$th row is
$Z_{i}=(Z_{i1},\ldots, Z_{ip})$, where $Z_{ij}$ is the $j$th
observation for subject $i$. Let $M=(M_{ij})$ be a missing data
indicator matrix with the $i$th row $M_{i}=(M_{i1},\ldots, M_{ip})$,
such that $M_{ij}$ is 1 if $Z_{ij}$ is missing and $M_{ij}$ is 0 if
$Z_{ij}$ is present. We assume that $(Z_{i},M_{i}),$ $i=1,\ldots,n$,
are independent and identically distributed. In Rubin (\citeyear{r33}),
the joint distribution is factored as
\begin{equation}\label{selection}
f({Z}_{i},{M}_{i}|\theta, \psi )=f({Z}_{i}|\theta
)f({M}_{i}|{Z}_{i},\psi ),
\end{equation}%
where $f(Z_{i}|\theta )$ represents the model for the data without
missing
values, $f(M_{i}|Z_{i},\psi )$ models the missing-data mechanism, and $%
(\theta, \psi )$ denotes unknown parameters. When missingness does not
depend on the values of the data $Z$, missing or observed, that is, if
\[
f(M_{i}|Z_{i},\psi )=f(M_{i}|\psi )\quad\mbox{for all }Z_{i},\psi,
\]
the data are called missing completely at random (MCAR). With the
exception of some planned miss\-ing-data designs, MCAR is a strong
assumption, and
missingness often depends on the observed and/or unobserved data. Let $Z_{%
\mathrm{obs},i}$ denote the observed component of $Z_{i}$ and $Z_{\mathrm{mis}%
,i}$ the missing component. A less restrictive assumption is that
missingness depends only on the observed values $Z_{\mathrm{obs},i}$,
and not on the missing values $Z_{\mathrm{mis},i}$. That is,
\[
f(M_{i}|Z_{i},\psi )=f(M_{i}|{Z_{\mathrm{obs},i}},\psi )\quad\mbox{for
all }{Z_{\mathrm{mis},i}},\psi .
\]
The missing-data mechanism is then called missing at random (MAR). The
mechanism is called missing not at random (MNAR) if the distribution of
$M$ depends on the missing values in the data matrix~$Z$.

The observed data consist of the values of the variables
$(Z_{\mathrm{obs}},M)$ and the distribution of the observed data
is obtained by integrating $Z_{\mathrm{mis}}$ out of the joint density of $%
Z=(Z_{\mathrm{obs}},Z_{\mathrm{mis}})$ and $M$. That is, for unit~$i$,

\begin{eqnarray}\label{likelihood}
&&f(Z_{\mathrm{obs},i},M_{i}|\theta, \psi )\nonumber
\\
&&\quad{}=\int f(Z_{\mathrm{obs},i},Z_{%
\mathrm{mis},i}|\theta )
\\
&&\qquad\hphantom{\int}{}\cdot
f(M_{i}|Z_{\mathrm{obs},i},Z_{\mathrm{mis},i},\psi
)\,dZ_{\mathrm{mis},i}.\nonumber
\end{eqnarray}
The full likelihood of $\theta $ and $\psi $ is any function of $\theta
$ and $\psi $ proportional to the product of (\ref{likelihood}) over
observations $i$:
\[
L_{\mathrm{full}}(\theta, \psi |Z_{\mathrm{obs}},M)\propto \prod_{i=1}^{n}f({Z_{%
\mathrm{obs},i}},M_{i}|\theta, \psi ).
\]
The missing-data mechanism is called ignorable if it is MAR and if in
addition, the parameter space for $(\theta, \psi )$ is a Cartesian
product space $\Theta \times \Psi $ where $\theta\in \Theta $ and
$\psi\in\Psi $. Likelihood-based inferences for $\theta $ can then be
based on
\[
L_{\mathrm{ign}}(\theta |Z_{\mathrm{obs}})\propto \prod_{i=1}^{n}f(Z_{\mathrm{obs}%
,i}|\theta ),
\]
the ignorable likelihood of $\theta$ based on the observed data $Z_{%
\mathrm{obs}}$ (Rubin, \citeyear{r33}). Many methods of handling
missing data assume missingness is MCAR or MAR. If this is assumed, the
missing-data mechanism can be ignored and we only need to model the
observed data $Z_{\mathrm{obs}}$ to derive likelihood-based inferences
for $\theta$. However, these inferences are subject to bias when the
data are not MAR.

Equation (\ref{selection}) is sometimes
called a selection model factorization of the joint distribution of $%
(Z_{i},M_{i})$ because of connections with the econometric literature
on selection bias (Heckman, \citeyear{r14}). Clearly other
factorizations are possible. In particular, pattern-mixture models
(Little, \citeyear{r19}) factor the joint distribution as
\begin{equation}\label{pattern-mixture}
f({Z}_{i},{M}_{i}|\varphi, \pi )=f({M}_{i}|\pi )f({Z}_{i}|{M}%
_{i},\varphi ),
\end{equation}%
which models the distribution of $Z_{i}$ for each pattern of missing
data.

Both selection and pattern-mixture models treat the variables ${Z}_{i}$
and missing-data indicators ${M}_{i}$ as single blocks. Little
attention has been paid to models that disaggregate these blocks based
on subsets of variables and their missing-data
indicators. One such class of models is generated by writing ${Z}_{i}=({Z}%
_{i(1)},{Z}_{i(2)},\ldots, {Z}_{i(B)})$ where ${Z}_{i(j)}$ is a subset
of
the variables, with corresponding missing-data indicators ${M}%
_{i}=(M_{i(1)},M_{i(2)},\ldots, M_{i(B)})$. For convenience, define the
``history'' up to block $j$ for unit $i$ as
\[
\mathcal{H}_{i(j)}=\bigl(Z_{i(1)},M_{i(1)},\ldots
,Z_{i(j)},M_{i(j)}\bigr)
\]
and factor the joint distribution as
\begin{eqnarray}\label{block-sequential}
&&\hspace*{15pt}f({Z}_{i},{M}_{i}|\theta, \psi )\nonumber \hspace*{-9pt}
\\
&&\hspace*{15pt}\quad= f\bigl({Z}_{i(1)},{M}_{i(1)}|{\theta }%
^{(1)},{\psi }^{(1)}\bigr)\nonumber\hspace*{-9pt}
\\[-8pt]\\[-8pt]
&&\hspace*{15pt}\qquad{}\cdot f\bigl({Z}_{i(2)},{M}_{i(2)}|\mathcal{H}_{i(1)},{%
\theta }^{(2)},{\psi }^{(2)}\bigr)  \nonumber\hspace*{-9pt}\\
&&\hspace*{15pt}\qquad{} \cdot \cdots \cdot f\bigl({Z}_{i(B)},{M}_{i(B)}|\mathcal{H}_{i(B-1)},
{\theta }%
^{(B)},{\psi }^{(B)}\bigr).\nonumber\hspace*{-9pt}
\end{eqnarray}
We call models based on the factorization (\ref{block-sequential}) \textit{%
block-sequential missing data models}. The set
$({Z}_{i(j)},{M}_{i(j)})$ in the $j$th block might be modeled using the
selection or pattern-mixture
factorization, yielding combinations of (\ref{selection}) and (\ref%
{pattern-mixture}). This approach to modeling might be seen as natural
when the blocks unfold sequentially in time, or if they follow a causal
sequence, and the variables in a block are conditioned on prior
variables in time or in the causal chain. Along these lines, Robins and
Gill (\citeyear{r31}) and Robins (\citeyear{r30}) argue that MAR is
hard to justify causally when data do not have a monotone pattern, and
discuss alternative factorizations that have a readier causal
interpretation.

Various modeling assumptions might be incorporated in
(\ref{block-sequential}). In this article we consider a particular form
of potentially MNAR models based on (\ref{block-sequential}) with
specific assumptions concerning the dependence of the distribution of
the variables in each block on the history. Specifically, we assume
that in
the $j$th block, the joint distribution of $({Z}_{i(j)},{M}_{i(j)}|\break\mathcal{H%
}_{i(j-1)}$) can be factorized as follows (parameters are left
implicit):
\begin{eqnarray}\label{BCMAR}
&&\hspace*{15pt}f\bigl({Z}_{i(j)},{M}_{i(j)}|\mathcal{H}_{i(j-1)}\bigr)\nonumber\hspace*{-15pt}
\\[-8pt]\\[-8pt]
&&\hspace*{15pt}\quad=f\bigl({Z}_{i(j)}|\mathcal{H}
_{i(j-1)}\bigr)
f\bigl({M}_{i(j)}|\mathcal{H}_{i(j-1)},Z_{i(j)}\bigr),\nonumber\hspace*{-15pt}
\end{eqnarray}
where
\begin{eqnarray*}
 f\bigl({Z}_{i(j)}|\mathcal{H}_{i(j-1)}\bigr)&=&f\bigl({Z}_{i(j)}|{Z}_{i(1)},\ldots, {Z}%
_{i(j-1)}\bigr),  \\
 f\bigl({M}_{i(j)}|\mathcal{H}_{i(j-1)},Z_{i(j)}\bigr)&=&f\bigl({M}_{i(j)}|\mathcal{H}%
_{i(j-1)},Z_{\mathrm{obs},i(j)}\bigr),
\end{eqnarray*}
and ${Z}_{\mathrm{obs},i(j)}$ denotes the observed components of ${Z}_{i(j)}$%
. That is, the distribution of $Z_{i(j)}$ given the previous variables
depends only on the previous $Z$'s, not the previous $%
M$'s, and the distribution of $M_{i(j)}$ can depend
on
previous $Z$'s, $M$'s and ${Z}_{\mathrm{obs}%
,i(j)}$, but not on the missing components of ${Z}_{i(j)}$, say, $Z_{%
\mathrm{mis},i(j)}$. We call models of the form (\ref{BCMAR}) \textit{%
block-conditional MAR \textup{(}BCMAR}), since each block would be MAR if values of $%
Z $ in previous blocks were fully observed.

For $B=2$ blocks, (\ref{BCMAR}) reduces to
\begin{eqnarray}\label{BCMAR2}
&&\quad f({Z}_{i},{M}_{i}|\theta, \psi )\nonumber
\\
&&\quad\quad= f\bigl({Z}_{i(1)}|{\theta }^{(1)}\bigr)f\bigl({M}_{i(1)}|
{Z}_{\mathrm{obs},i(1)},{\psi }^{(1)}\bigr) \nonumber
\\[-8pt]\\[-8pt]
&&\quad\qquad{}\cdot f\bigl({Z}_{i(2)}|{Z}_{i(1)},{\theta
}^{(2)}\bigr)\nonumber
\\
&&\quad\qquad{}\cdot f\bigl({M}_{i(2)}|{M}_{i(1)},{Z}%
_{i(1)},{Z}_{\mathrm{obs},i(2)},{\psi }^{(2)}\bigr),\nonumber
\end{eqnarray}
where $Z_{i(1)}$ is MAR, ignoring information about $Z_{i(2)}$ and $M_{i(2)}$%
, and missingness of $Z_{i(2)}$ depends on the observed components of $%
Z_{i(2)}$, observed and unobserved value of $Z_{i(1)}$ and on
$M_{i(1)}$. This mechanism is not in general MAR, since missingness of
$Z_{i(2)}$ is allowed to depend on missing values of
$Z_{\mathrm{mis},i(1)}$. For the particular case where ${Z}_{i(1)}$ and
${Z}_{i(2)}$ are single variables, this reduces to the simpler form
\begin{eqnarray}\label{BCMAR2A}
&&f({Z}_{i},{M}_{i}|\theta, \psi )\nonumber
\\
&&\quad=f\bigl({Z}_{i(1)}|{\theta}^{(1)}\bigr)f\bigl({M}_{i(1)}|{%
\psi }^{(1)}\bigr)\nonumber
\\[-8pt]\\[-8pt]
&&\qquad{}\cdot f\bigl({Z}_{i(2)}|{Z}_{i(1)},{\theta }^{(2)}\bigr)\nonumber
\\
&&\qquad{}\cdot f\bigl({M}_{i(2)}|{M}%
_{i(1)},{Z}_{i(1)},{\psi }^{(2)}\bigr),\nonumber
\end{eqnarray}
because of the MAR condition in each block. In this case, $Z_{i(1)}$ is
MCAR and, given $Z_{i(1)},M_{i(1)},Z_{i(2)}$ is also MAR. In Section \ref{2}
we describe inference for BCMAR models based on a block-monotone
reduced likelihood, where the conditional distribution of the variables
in each block, given the variables in previous blocks, is computed
using only the subset of cases for which the variables in previous
blocks are fully observed. This reduced likelihood is related but not
quite the same as a partial likelihood as defined by Cox
(\citeyear{r6}). This reduced likelihood does not require a model for
the distribution of the missing-data indicators $M$. This is a useful
property, since specifying models for $M$ can be challenging, and
results are vulnerable to misspecification. The block-monotone reduced
likelihood becomes the full likelihood when data have a particular
pattern, which we call \textit{block monotone}.

Use of the block-monotone reduced likelihood generally involves a loss
of information, and an interesting question is how much information is
lost; the remainder of the paper examines this question in the context
of simple bivariate examples. We analyze in detail the model
(\ref{BCMAR2A}) for case of bivariate categorical $Z$, where the
complete cases form a 2-way contingency table, and the incomplete cases
form supplemental margins (see, for example, Little and Rubin,
\citeyear{r22}, Chapter 13). In addition, we give a less detailed
analysis of a more general example with two blocks where the
distribution of $Z_{i(2)}$ is from the exponential family.

The EM algorithm (Dempster, Laird and Rubin, \citeyear{r7}), a
ubiquitous algorithm for ML estimation from incomplete data and the
topic of this special issue, plays a useful role in fitting these
models. EM is particularly appealing for categorical data, since the
Poisson and multinomial distributions for modeling count data yield
complete data loglikelihoods that are linear in the cell counts.
Consequently, the E step of EM consists of replacing the complete-data
cell counts by conditional expectations given the observed data, in
effect distributing the supplemental margins into the full table
according to current estimates of the cell probabilities. The M step of
EM is the same as complete-data ML estimation based on the data filled
in by the E step. This approach to estimation for count data with some
grouped counts was first established as ML by Hartley (\citeyear{r13}).
The application to a ($2\times2$) table with supplemental margins was
considered by Chen and Fienberg (\citeyear{r5}), and extended to the
general class of loglinear models by Fuchs (\citeyear{r11}).

For some hierarchical loglinear models the M step of EM requires
iteration, so EM involves double iteration. The usual approach is the
Deming--Stephan algorithm, also known as iterative proportional fitting
(Bishop, Fienberg and Holland, \citeyear{r4}). If the M step is
restricted to just one iteration of Deming--Stephan, the result is an
example of an ECM\break (Expectation Conditional Maximization) algorithm,
which achieves similar theoretical properties to EM with just a single
iterative loop (Meng and Rubin, \citeyear{r25}; Little and Rubin,
\citeyear{r22}). EM is also useful for fitting MNAR models for
contingency tables (Baker and Laird, \citeyear{r2}; Fay, \citeyear{r9};
Rubin, Stern and Vehovar, \citeyear{r35}; Little and Rubin
\citeyear{r22}, Section 15.7). As shown below, EM also plays a useful
role for \mbox{BCMAR} models.

In Section \ref{3}, we consider ML estimation for a \mbox{BCMAR} model for bivariate
categorical data, where $Z=(Z_{(1)},Z_{(2)})$ are assumed to have a
multinomial distribution. The results are surprising. The
block-monotone reduced ML estimates of the parameters of the joint
distribution of $(Z_{(1)},Z_{(2)})$ (as discussed in Section \ref{2}) are
computed noniteratively from the monotone pattern, excluding the data
with $Z_{(2)}$ observed and $Z_{(1)}$ missing. These are in fact the
full ML estimates, providing corresponding estimates of the parameters
of the missing-data mechanism all lie in the admissible range $[0,1]$.
If not, then the data with $Z_{(2)}$ observed and $Z_{(1)}$ missing
enter into the full ML estimates, and an iterative algorithm such as EM
is needed to compute them. In Section \ref{4}, a restricted version of the
\mbox{BCMAR} model is introduced where missingness of $Z_{(2)}$ depends on the
perhaps unobserved value of $Z_{(1)}$ but not on whether $Z_{(1)}$ is
missing. Some numerical examples are presented in Section \ref{5} to compare
unrestricted and restricted BCMAR models and MAR models and to
illustrate when the block-monotone reduced ML estimates in the BCMAR
models are full ML. A~real data example is given in Section \ref{6}. Section~\ref{7}
explores a more general example of a BCMAR model with two blocks, in
which the possibly vector valued variable $Z_{(2)}$ arises from a
distribution in the exponential family. Section \ref{8} reviews the ideas of
the article and outlines extensions to other missing-data problems.

\section{Estimation of Block-Conditional MAR Models Using a Reduced
Likelihood}\label{2}

For any BCMAR model, define the \textit{%
block-monotone reduced likelihood} to be
\begin{eqnarray}\label{partial}
&&{L}_{\mathrm{bm}}(\theta )\nonumber
\\
&&\quad=\prod_{j=1}^{B}\prod_{i\in Q_{j}}f\bigl({Z}_{%
\mathrm{obs},i(j)}|{Z}_{i(1)},{Z}_{i(2)},\ldots,
\\
&&\hspace*{127pt}{Z}_{i(j-1)},{\theta }%
^{(j)}\bigr),\nonumber
\end{eqnarray}
where ${Q}_{j}$ is the subset of
cases with ${Z}_{i(1)},{Z}_{i(2)},\ldots, \break{Z}_{i(j-1)}$ fully observed, that is, $%
M_{i(1)}=M_{i(2)}=\cdots=M_{i(j-1)}=0$. Under usual regularity
conditions, the estimator of $\theta$ that maximizes ${L}%
_{\mathrm{{bm}}}(\theta )$ has the same properties as maximum
likelihood, in that it is consistent and asymptotically normal with an
asymptotic covariance matrix estimated by $I(\hat\theta)^{-1}$ where
$I(\theta)=-\partial^2 \log L_{\mathrm{bm}}(\theta)/\partial
\theta^T\,\partial \theta$. These results can be obtained using
conditional arguments similar to those of Cox (\citeyear{r6}) in his
examination of partial likelihood.

We prove this property for the special case of $B=2$ blocks; the
extension to more than two blocks is straightforward. The observed-data
likelihood for the two blocks can be written
\begin{eqnarray}\label{partial2}
 &&
 \hspace*{8pt}{L}_{\mathrm{obs}}(\theta,\psi )\nonumber\hspace*{-8pt}
 \\
 &&\hspace*{8pt}\quad=
 \prod_{i=1}^{n}\bigl\{f\bigl({Z}_{\mathrm{obs}%
,i(1)},M_{i(1)}|\theta, \psi\bigr)\nonumber\hspace*{-9pt}
\\
&&\hspace*{8pt}\hphantom{\prod_{i=1}^{n}\bigl\{}\qquad{}\cdot
\bigl[f\bigl({Z}_{\mathrm{obs},i(2)},M_{i(2)}|{Z}_{\mathrm{obs},i(1)},\nonumber\hspace*{-8pt}
\\[-8pt]\\[-8pt]
&&\hspace*{8pt}\qquad\quad{}\hphantom{\prod_{i=1}^{n}\bigl\{\times
\bigl[f\bigl({Z}_{\mathrm{obs},i(2)},M_{5}}
M_{i(1)}=0,\theta
,\psi\bigr)\bigr]^{\delta_i}\nonumber\hspace*{-8pt}
\\
&&\hspace*{8pt}\hphantom{\prod_{i=1}^{n}\bigl\{}\qquad{}\cdot
\bigl[f\bigl({Z}_{\mathrm{obs},i(2)},M_{i(2)}|{Z}_{\mathrm{obs},i(1)},\nonumber\hspace*{-8pt}
\\
&&\hspace*{8pt}\qquad\quad{}\hphantom{\prod_{i=1}^{n}\bigl\{\cdot\,\,
\bigl[f\bigl({Z}_{\mathrm{obs},i(2)},M_{i(2)}}
M_{i(1)},\theta,\psi\bigr)\bigr]^{1-\delta_i}\bigr\},\nonumber\hspace*{-8pt}
\end{eqnarray}
where $\delta_i=I(M_{i(1)}=0)$. Note that the second term in the
product refers to the cases for which $i\in Q_2$. Consider the pseudo-likelihood
generated by the first two terms in the product
(\ref{partial2}). Let $\gamma=(\theta,\psi)$,
and denote the corresponding scores as
\[
S_{i(1)}=\frac{\partial}{\partial \gamma} \log f\bigl({Z}_{\mathrm{obs}%
,i(1)},M_{i(1)}|\theta, \psi \bigr)
\]
and
\begin{eqnarray*}
S_{i(2)}&=&\delta_i
\frac{\partial}{\partial \gamma}
\\
&&{}\cdot\log
f\bigl({Z}_{\mathrm{obs},i(2)},M_{i(2)}|{Z}_{\mathrm{obs},i(1)},
\\
&&\hphantom{\cdot\log
f\bigl({Z}_{\mathrm{obs},i(2)},M_{i(2)}2}{}M_{i(1)}=0,\theta, \psi \bigr).
\end{eqnarray*}
Under usual regularity
conditions for the appropriate conditional densities, it is now easily
seen that $E[S_{i(j)}]=0$ and $E[S_{i(j)}^2]=-E[\partial
S_{i(j)}/\partial \gamma]$ where $j=1,2$. Finally, by conditioning on
$Z_{\mathrm{obs},i(1)},M_{i(1)}$, it can be seen that
$E[S_{i(1)}S_{i(2)}]=0$ so that the scores are uncorrelated. It follows
that
\begin{equation}\label{esteq}
\sum_{i=1}^n \bigl[S_{i(1)}(\theta,\psi)+S_{i(2)}(\theta,\psi)\bigr]=0
\end{equation}
is an unbiased estimating equation with asymptotic properties similar
to those of a likelihood score equation. Under i.i.d. assumptions for the
data $\{(Z_{i(1)},\break M_{i(1)}, Z_{i(2)}, M_{i(2)}),i=1,\ldots,n\}$, the
central limit theorem applies to the total score and a Taylor expansion
gives the usual asymptotic normal results for the estimators
$\hat\theta, \hat\psi$ that arise as a solution to (\ref{esteq}).
Further, the asymptotic variance of $\hat\theta, \hat\psi$ can be
estimated as the inverse of the usual observed information. Finally, we
note that
\begin{eqnarray*}
&&
\hspace*{-5pt}{L}_{\mathrm{obs}}(\theta, \psi )
\\
&&\hspace*{-5pt}\quad=
\prod_{i=1}^{n}f\bigl({Z}_{\mathrm{obs},i(1)}|{\theta }^{(1)}\bigr)f\bigl({M}_{i(1)}|
Z_{\mathrm{obs},i(1)},{\psi }^{(1)}\bigr)
\\
&&\hspace*{-5pt}\qquad{}\cdot
\prod_{i\in Q_{2}}f\bigl({Z}_{\mathrm{obs},i(2)}|{Z}_{i(1)},{\theta }^{(2)}\bigr)
\\
&&\hspace*{-5pt}\qquad\hphantom{\times\prod_{i\in Q_{2}}}{}\cdot f\bigl({M}_{i(2)}|{Z}_{i(1)},
{M}_{i(1)}=0,Z{_{\mathrm{obs},i(2)}},{\psi }^{(2)}\bigr)
\\
&&\hspace*{-5pt}\qquad{}\cdot \prod_{i\notin Q_{2}}f\bigl({Z}_{\mathrm{obs},i(2)},M_{i(2)}|
{Z}_{\mathrm{obs},i(1)},M_{i(1)},\theta, \psi \bigr),
\end{eqnarray*}
where the factorization of the first two products into distinct
components for $\theta$ and $\psi$ is a result of the \mbox{BCMAR}
assumptions. Rearranging terms, we can write
\[
{L}_{\mathrm{obs}}(\theta, \psi )={L}_{\mathrm{bm}}(\theta )\times {L}_{\mathrm{M}}
(\psi )\times {L}_{\mathrm{rest}}(\theta, \psi ),
\]
where
\begin{eqnarray*}
 {L}_{\mathrm{bm}}(\theta )
 &=&
\prod_{i=1}^{n}f\bigl({Z}_{\mathrm{obs},i(1)}|{\theta }^{(1)}\bigr),
\\
&&{}\cdot
\prod_{i\in Q_{2}}f\bigl({Z}_{\mathrm{obs},i(2)}|{Z}_{i(1)},{\theta}^{(2)}\bigr)
\\
{L}_{\mathrm{M}}(\psi )
&=&
\prod_{i=1}^{n}f\bigl({M}_{i(1)}|{Z}_{\mathrm{obs},i(1)},{\psi }^{(1)}\bigr)
\\
&&{}
\cdot \prod_{i\in Q_{2}}
f\bigl({M}_{i(2)}|{Z}_{i(1)},{M}_{i(1)}=0,
\\
&&\hphantom{\,|{Z}_{i(1)},{M}_{i(1)}=0,}
{Z}_{\mathrm{obs},i(2)},{\psi }^{(2)}\bigr),
\\
 {L}_{\mathrm{rest}}(\theta, \psi)
 &=&
 \prod_{i\notin Q_{2}}f\bigl({Z}_{\mathrm{obs},i(2)},M_{i(2)}|{Z}_{\mathrm{obs},i(1)},
 \\
 &&\hphantom{\prod_{i\notin Q_{2}}f\bigl({Z}_{\mathrm{obs},i(2)},M_{i(2)}|}
 M_{i(1)},\theta, \psi\bigr).
\end{eqnarray*}
It can then be easily seen that the observed information matrix based on the first two
components is
diagonal in the parameters, and the asymptotic results for $\theta$ can
be determined from ${L}_{\mathrm{bm}}(\theta)$ as described above.

The block-monotone reduced likelihood inference drops the components
${L}_{\mathrm{M}}(\psi)$ and ${L}_{\mathrm{rest}}(\theta, \psi)$ from
the likelihood, and bases inference about $\theta$ on the remaining
term ${L}_{\mathrm{bm}}(\theta).$ This provides a convenient approach
to inference, since the block-monotone reduced likelihood does not
involve the distributions of the missing-data indicators, and, hence,
these distributions do not need to be specified. Correctly specifying
these distributions is not easy, and estimates of $\theta$ are
vulnerable to their misspecification.

We say that ${Z}_{i}=({Z}_{i(1)},{Z}_{i(2)},\ldots, {Z}_{i(B)})$ have a
\textit{block monotone} pattern if, for all $j$, $Z_{i(j-1)}$ is fully
observed whenever $Z_{i(j)}$ has at least one observed component. Note
that block monotonicity is weaker than a monotone pattern for all the
variables, since the variables within each block do not necessarily
have a monotone pattern. If the data have a block monotone pattern, the
term ${L}_{\mathrm{rest}}(\theta,\psi)$ is no longer present, and the
block-monotone reduced likelihood is equivalent to the full likelihood
for inference about $\theta$, providing the parameters $\theta$
and $\psi$ are distinct. In other situations, dropping the term ${L}_{%
\mathrm{rest}}(\theta,\psi)$ involves a loss of information, so the
estimates are not in general fully efficient compared with full ML. We
explore this potential loss in efficiency for some simple models in the
remainder of this article.

\section{Unrestricted BCMAR Models for Bivariate Categorical
Data}\label{3}

We consider data with $B=2$, $Z=(Z_{(1)},Z_{(2)})$ whe\-re $Z_{(1)}$ and
$Z_{(2)}$ are categorical variables with $J$ and $K$ categories
respectively. Both $Z_{(1)}$ and $Z_{(2)}$ may be missing, so there are
four missing-data patterns. Let $r=0,1,2,3$ index the missing-data
patterns and let $P_{r}$ denote the set of sample cases with pattern
type $r,r=0,\ldots,3$ (see Table \ref{t1}). Let $n_{r}$ denote the number of
cases in the sample with pattern $r$ and $n=\sum_{r}n_{r}$ denote the
total sample size.

\begin{table}[b]
\caption{Missing-data pattern for two variables}\label{t1}
\begin{tabular*}{50pt}{@{\extracolsep{\fill}}l|>{\columncolor{black}\vline}c|>
{\columncolor{black}\vline}c|@{}}
 \multicolumn{1}{@{}l}{Pattern}&\multicolumn{2}{l@{}}{}\\
$P_0$&&\\
 $P_1$ && \multicolumn{1}{>{\columncolor{white}}c|}{?}\\
 $P_2$ & \multicolumn{1}{>{\columncolor{white}}c|}{?} &\\[-1pt]
\hhline{~--} $P_3$ & \multicolumn{1}{>{\columncolor{white}}c|}{?}
&\multicolumn{1}{>{\columncolor{white}}c|}{?}\\[-1pt]
\hhline{~--}
\end{tabular*}
\end{table}

For categorical $Z_{(1)}$ and $Z_{(2)}$ with $J$ and $K$ levels, data
in $P_{0}$ can be arranged as a $J\times K$ contingency table, and the
data in $P_{1}$ and $P_{2}$ form supplemental $J\times 1$ and $1\times
K$ margins. Let $n_{(0),jk}$ be the count of complete cases with
$Z_{(1)}=j,Z_{(2)}=k$, $n_{(1),j+}$ be the count of cases with
$Z_{(1)}=j$ and $Z_{(2)}$ missing, $n_{(2),+k}$ be the count of cases
with $Z_{(2)}=k$ and $Z_{(1)}$ missing, and $n_{(3),++}$ be the count
of cases with both $Z_{(1)}$ and $Z_{(2)}$ missing. The data are
displayed in Table \ref{t2}. Note that
$n_{0}=\sum_{j=1}^{J}\sum_{k=1}^{K}n_{(0),jk}$,
$n_{1}=\sum_{j=1}^{J}n_{(1),j+}$, $n_{2}=\sum_{k=1}^{K}n_{(2),+k}$, and
$n_{3}=n_{(3),++}$.

The parameters of interest are $\theta=\{\theta _{jk}\}$, where $\theta
_{jk}=P(Z_{(1)}=j,Z_{(2)}=k)$ with $\sum_{j=1}^{J}\sum_{k=1}^{K}\theta
_{jk}=1$. The MAR assumption for these data implies that
\begin{eqnarray*}
P\bigl(M_{(1)}=M_{(2)}=1|Z_{(1)}=j,Z_{(2)}=k\bigr)&=&\upsilon,  \\
P\bigl(M_{(1)}=0,M_{(2)}=1|Z_{(1)}=j,Z_{(2)}=k\bigr)&=&\upsilon _{j}^{(0)},  \\
P\bigl(M_{(1)}=1,M_{(2)}=0|Z_{(1)}=j,Z_{(2)}=k\bigr)&=&\upsilon _{k}^{(1)},
\end{eqnarray*}\vspace*{-20pt}
\begin{eqnarray*}
&&P\bigl(M_{(1)}=M_{(2)}=0|Z_{(1)}=j,Z_{(2)}=k\bigr)
\\
&&\quad=
1-\upsilon-\upsilon_{j}^{(0)}-\upsilon_{k}^{(1)},
\end{eqnarray*}
where $1\leq j\leq J,1\leq k\leq K$ and $M_{(1)}$ and
$M_{(2)}$ are missing-data indicators for $Z_{(1)}$ and $Z_{(2)}$ with
$1$ and~$0$ denoting missing and observed values respectively (see
Little and Rubin, \citeyear{r22}, Example 1.19). In this case, $\zeta
=\{\upsilon, \upsilon _{j}^{(0)},\upsilon _{k}^{(1)}\}$ represent
nuisance parameters for the missing-data mechanism. Under MAR, the
likelihood factors into distinct components of $\theta$ and $\zeta$; ML
estimation of $\theta$ under MAR involves all the observed data and
typically requires an iterative algorithm such as EM (Little and Rubin,
\citeyear{r22}, Chapter 13).

\begin{table}
\tabcolsep=0pt
\caption{Notation for a $J\times K$ table with supplemental
margins for both variables}\label{t2}
\begin{tabular*}{\columnwidth}{@{\extracolsep{\fill}}cccccccc@{}}
\hline
& & \multicolumn{6}{c@{}}{$\bolds{Z_{(2)}}$} \\
\cline{3-8}\\ [-5pt]
& & \textbf{1} & \textbf{2 }& \textbf{\ldots} & \textbf{\ldots} & $\bolds{K}$ &
\textbf{Missing} \\
\hline
& \textbf{1} & $n_{(0),11}$ & $n_{(0),12}$ & \ldots & \ldots & $n_{(0),1K}$ & $n_{(1),1+}$\\
& \textbf{2} & $n_{(0),21}$ & $n_{(0),22}$ & \ldots & \ldots & $n_{(0),2K}$ &$n_{(1),2+}$\\
$\bolds{Z_{(1)}}$ & $\bolds{\vdots}$ & \vdots & \vdots & \vdots & \vdots & \vdots & \vdots \\
& \textbf{J} & $n_{(0),J1}$ & $n_{(0),J2}$ & \ldots & \ldots & $n_{(0),JK}$ &$n_{(1),J+}$\\
\\[3pt]
& \textbf{Missing} & $n_{(2),+1}$ & $n_{(2),+2}$ & \ldots & \ldots & $n_{(2),+K}$ & $n_{(3),++}$\\
\hline
\end{tabular*}
\end{table}

We consider as an alternative to MAR
the following BCMAR model (\ref{BCMAR2A}), which incorporates
the
assumption that $Z_{(1)}$ is MCAR and missingness of $Z_{(2)}$ depends on $%
Z_{(1)}$ and $M_{(1)}$:
\begin{eqnarray}\label{E:mechanism}
&& P\bigl(M_{(1)}=1|Z_{(1)}=j,Z_{(2)}=k\bigr)=\phi,  \nonumber
\\
&& P\bigl(M_{(2)}=1|M_{(1)}=0,Z_{(1)}=j,Z_{(2)}=k\bigr)\nonumber
\\
&&\quad=\phi _{j}^{(0)},
 \\
&& P\bigl(M_{(2)}=1|M_{(1)}=1,Z_{(1)}=j,Z_{(2)}=k\bigr)\nonumber
\\
&&\quad=\phi _{j}^{(1)},\nonumber
\end{eqnarray}
where $1\leq j\leq J,1\leq k\leq K$. Here $\Phi
=\{\phi, \phi _{j}^{(0)},\phi _{j}^{(1)}\}$ are nuisance parameters
corresponding to the missing-data mechanism. The number of parameters
in this model is $JK+2J$, whereas the degrees of freedom of the data
are $JK+J+K $, which comprise $JK$ for the complete cases, plus $J$ for
the supplemental margin on $Z_{(1)}$, plus $K$ for the supplemental
margin on $Z_{(2)}$, plus $1$ for the number of cases with $Z_{(1)}$
and $Z_{(2)}$ both missing, minus 1 for the total which is considered
fixed at $n$. When $J=K$, the model has the same number of parameters
as degrees of freedom in the data; otherwise, the model has more
parameters for $J>K$ or fewer for $J<K$.

Note that if $\phi _{j}^{(1)}=\phi^{(1)}$
does not depend on $j$, this reduces to a restricted MAR model in which $%
Z_{(1)}$ is MCAR and missingness of $Z_{(2)}$ depends on $M_{(1)}$, and
only depends on $Z_{(1)}$ for the pattern with $Z_{(1)}$ observed. A
likelihood ratio test could be used to test this restricted MAR
assumption against the more general BCMAR model and the EM algorithm
can be applied to compute the ML estimates (Little and Rubin,
\citeyear{r22}, Chapter 13). This restricted MAR model is introduced as
a testable submodel of the unrestricted BCMAR model, but we do not view
it as particularly appealing substantively, since if missingness of
$Z_{(2)}$ depends on $Z_{(1)}$ for the cases with $Z_{(1)}$ observed,
one might also expect it to depend on $Z_{(1)}$ for the cases with
$Z_{(1)}$ missing. Another submodel of the unrestricted BCMAR model is
discussed in Section \ref{4}.

\subsection{EM Algorithm}\label{31}

The full likelihood for the above model is
\begin{eqnarray}\label{full}
&&
L\bigl(\theta, \Phi|Z_{\mathrm{obs},(1)},Z_{\mathrm{obs},(2)},M\bigr) \nonumber
\\
&&\quad=
\prod_{i\in P_{0}}p\bigl(Z_{i(1)},Z_{i(2)}|\theta\bigr)(1-\phi)\nonumber
\\
&&\qquad\hphantom{\prod_{i\in P_{0}}}{}\cdot
p\bigl(M_{i(2)}=0|Z_{i(1)},M_{i(1)}=0,\Phi\bigr) \nonumber
\\
&&\qquad{}\cdot
\prod_{i\in P_{1}}p\bigl(Z_{i(1)}|\theta\bigr)(1-\phi)\nonumber
\\
&&\qquad\hphantom{\quad\prod_{i\in P_{1}}}{}\cdot
p\bigl(M_{i(2)}=1|Z_{i(1)},M_{i(1)}=0, \Phi\bigr)
\\
&&\qquad{}\cdot
\prod_{i\in P_{2}}\sum_{Z_{i(1)}}p\bigl(Z_{i(1)},Z_{i(2)}|\theta\bigr)\phi \nonumber
\\
&&\qquad\hphantom{\quad\prod_{i\in P_{2}}\sum_{Z_{i(1)}}}{}\cdot
p\bigl(M_{i(2)}=0|Z_{i(1)},M_{i(1)}=1,\Phi\bigr) \nonumber
\\
&&\qquad{}\cdot
\prod_{i\in P_{3}}\sum_{Z_{i(1)}}p\bigl(Z_{i(1)}|\theta\bigr)\phi \nonumber
\\
&&\qquad\hphantom{\quad\prod_{i\in P_{2}}\sum_{Z_{i(1)}}}{}\cdot
p\bigl(M_{i(2)}=1|Z_{i(1)},M_{i(1)}=1,\Phi\bigr).\nonumber
\end{eqnarray}
The block-monotone reduced likelihood
is
\begin{eqnarray} \label{reduced}
&&\quad L_{\mathrm{bm}}\bigl(\theta|Z_{\mathrm{obs},(1)},Z_{\mathrm{obs},(2)}\bigr)\nonumber
\\[-8pt]\\[-8pt]
&&\quad\quad=\prod_{i\in P_{0}}p\bigl(Z_{i(1)},Z_{i(2)}|\theta \bigr)
\prod_{i\in P_{1}}p\bigl(Z_{i(1)}|\theta\bigr),\nonumber
\end{eqnarray}
which does not model the missing data mechanism, and drops the data for
patterns $P_{2}$ and $P_{3}$. We first consider ML estimation for the
full likelihood (\ref{full}), and then discuss the relationship between
these ML estimates and the estimates that maximize the block-monotone
reduced likelihood (\ref{reduced}).

One approach to ML estimation is to
apply the EM algorithm. To define the E step of EM, let
$(\theta_{jk}^{(t)},\break
{\phi_{j}^{(1)}}^{(t)})$ denote the parameter estimates at iteration $t$, and $%
n_{(r),jk}^{(t)}$ be the estimate of cell frequency for $%
Z_{i(1)}=j,Z_{i(2)}=k$ in pattern $P_{r}$. The E step distributes the
partially classified observations into the table according to the
corresponding probabilities:
\begin{eqnarray*}
n_{(1),jk}^{(t)}&=& n_{(1),j+}\cdot \frac{\theta _{jk}^{(t)}}{%
\theta _{j+}^{(t)}}, \\
n_{(2),jk}^{(t)}&=& n_{(2),+k}\cdot \frac{(1-{\phi _{j}^{(1)}}%
^{(t)})\theta _{jk}^{(t)}}{\sum_{j=1}^{J}(1-{\phi
_{j}^{(1)}}^{(t)})\theta
_{jk}^{(t)}}, \\
n_{(3),jk}^{(t)}&=& n_{(3),++}\cdot \frac{{\phi _{j}^{(1)}}%
^{(t)}\theta _{jk}^{(t)}}{\sum_{j=1}^{J}{\phi _{j}^{(1)}}^{(t)}\theta
_{j+}^{(t)}}.
\end{eqnarray*}
The M step calculates new parameters as follows:
\begin{eqnarray*}
\theta_{jk}^{(t+1)}&=& \frac{n_{(0),jk} + n_{(1),jk}^{(t)} +
n_{(2),jk}^{(t)} + n_{(3),jk}^{(t)}}{n}, \\
\phi&=&\frac{\sum_{i=1}^{n}I(M_{i(1)}=1)}{n}=\frac{n_{2}+n_{3}}{n}, \\
{\phi_{j}^{(0)}}&=&\frac{\sum_{i=1}^{n}I(M_{i(1)}=0,M_{i(2)}=1,Z_{i(1)}=j)}
{\sum_{i=1}^{n}I(M_{i(1)}=0,Z_{i(1)}=j)}\\
&=&\frac{n_{(1),j+}}{n_{(1),j+}+n_{(0),j+}}, \\
{\phi_{j}^{(1)}}^{(t+1)}&=& \frac{\sum_{k}n_{(3),jk}^{(t)}} {
\sum_{k}n_{(2),jk}^{(t)}+\sum_{k}n_{(3),jk}^{(t)}} .
\end{eqnarray*}
The E step and M step alternate until the parameter estimates converge.

Note that $\phi$ and $\{\phi_{j}^{(0)}\}$ are estimated directly and
are unchanged throughout the EM algorithm. Com\-plete-case estimates or
estimates arising from the monotone pattern $P_{0} $ and $P_{1}$ can be
chosen as the starting values of $\{\theta_{jk}\} $, and the estimates
of $\{\phi _{j}^{(0)}\}$ or any constant in $(0,1)$ can be taken as
initial values of $\{\phi _{j}^{(1)}\}$. When $J>K$, the model has more
parameters than degrees of the freedom. In this case, multiple maxima
may exist, and depending on starting values, the EM algorithm can
converge to different estimates. This case will be discussed further
below.

\subsection{Noniterative ML Estimates}\label{32}

When $J\geq K$, noniterative estimates of the parameters can sometimes
be obtained using the factored likelihood method (Little and Rubin,
\citeyear{r22}, Chapter 7). We transform the parameters
$(\theta_{jk},\phi,\break\phi _{j}^{(0)},\phi _{j}^{(1)})$ to
\begin{eqnarray}\label{newparam}
\quad\alpha _{(0),jk}
&=&
P\bigl(Z_{(1)}=j,Z_{(2)}=k|M_{(1)}=M_{(2)}=0\bigr),\nonumber\hspace*{-10pt}
\\
\beta _{(1),j+}
&=&
P\bigl(Z_{(1)}=j|M_{(1)}=0,M_{(2)}=1\bigr),\nonumber
\\
\gamma _{(2),+k}
&=&
P\bigl(Z_{(2)}=k|M_{(1)}=1,M_{(2)}=0\bigr),\nonumber
\\
\pi _{0}
&=&
P\bigl(M_{(1)}=0,M_{(2)}=0\bigr),
\\
\pi _{1}
&=&
P\bigl(M_{(1)}=0,M_{(2)}=1\bigr),\nonumber
\\
\pi _{2}
&=&
P\bigl(M_{(1)}=1,M_{(2)}=0\bigr),\nonumber
\\
\pi_{3}
&=&
P\bigl(M_{(1)}=1,M_{(2)}=1\bigr),\nonumber
\end{eqnarray}
where $1\leq j\leq J,1\leq k\leq K$ and the following constraints
apply:
\begin{eqnarray*}
\sum_{j=1}^{J}\sum_{k=1}^{K}{\alpha _{(0),jk}}&=&1,\qquad\sum_{j=1}^{J}{\beta
_{(1),j+}}=1,
\\
\sum_{k=1}^{K}{\gamma _{(2),+k}}&=&1,\qquad\sum_{r=0}^{3}\pi
_{r}=1.
\end{eqnarray*}
These parameters correspond to a pattern-mixture factorization, as in (\ref%
{pattern-mixture}). The components of $(\theta, \Phi )=(\theta
_{jk},\phi, \phi _{j}^{(0)},\phi _{j}^{(1)})$ can be expressed in terms
of the new parametrization (\ref{newparam}) as follows:
\begin{eqnarray}\label{transform}
 \qquad\theta _{jk}
 &=&
 \biggl( \frac{\alpha _{(0),jk}}{\alpha _{(0),j+}}\biggr)
 \biggl( \frac{\pi_{0}\alpha_{(0),j+}+\pi _{1}\beta _{(1),j+}}{\pi _{0}+\pi_{1}}\biggr),
 \nonumber\hspace*{-15pt}
 \\
 \phi
 &=&
 1-\pi _{0}-\pi _{1},
 \\
 \phi _{j}^{(0)}
 &=&
 \frac{\pi _{1}\beta _{(1),j+}}{\pi _{0}\alpha_{(0),j+}+\pi _{1}\beta _{(1),j+}},\nonumber
\end{eqnarray}
and \{$\phi _{j}^{(1)},j=1,\ldots,J$\} is a solution to the $K$
simultaneous equations
\begin{eqnarray*}
\sum_{j=1}^{J}\bigl(1-\phi _{j}^{(1)}\bigr)\theta
_{jk}&=&P\bigl(M_{(2)}=0,Z_{(2)}=k|M_{(1)}=1\bigr)
\\
&=&\frac{\pi _{2}}{1-\pi _{0}-\pi _{1}}\gamma _{(2),+k},
\end{eqnarray*}
where $\alpha _{(0),j+}=\sum_{k=1}^{K}\alpha _{(0),jk}$.

Letting $(\varphi, \pi)$ represent
the parameters in (\ref{newparam}), the likelihood can be written as
\begin{eqnarray*}\label{E:factor}
&&
L\bigl(\varphi,\pi|Z_{\mathrm{obs},(1)},Z_{\mathrm{obs},(2)},M\bigr)
\\
&&\quad=
\prod_{i=1}^{n}p\bigl(M_{i(1)},M_{i(2)}\bigr)
\\
&&\qquad{}\cdot
\prod_{i\in p_{0}}p\bigl(Z_{i(1)},Z_{i(2)}|M_{i(1)}=0,M_{i(2)}=0\bigr)
\\
&&\qquad{}\cdot
\prod_{i\in p_{1}}p\bigl(Z_{i(1)}|M_{i(1)}=0,M_{i(2)}=1\bigr)
\\
&&\qquad{}\cdot
\prod_{i\in p_{2}}p\bigl(Z_{i(2)}|M_{i(1)}=1,M_{i(2)}=0\bigr)
\\
&&\quad=
\prod_{r=0}^{3}\pi_{r}^{n_{r}}\prod_{j,k=1}^{J,K}\alpha_{(0),jk}^{n_{(0),jk}}
\prod_{j=1}^{J}\beta_{(1),j+}^{n_{(1),j+}}
\prod_{k=1}^{K}\gamma _{(2),+k}^{n_{(2),+k}} .
\end{eqnarray*}
Maximizing the four terms in this likelihood yields
\begin{eqnarray*}
\hat{\alpha}_{(0),jk}&=&\frac{n_{(0),jk}}{n_{0}}, \qquad\hat{\beta}_{(1),j+}=
\frac{n_{(1),j+}}{n_{1}},
\\
 \hat{\gamma}_{(2),+k}&=&\frac{n_{(2),+k}}{%
n_{2}},\qquad \hat{\pi}_{r}=\frac{n_{r}}{n},
\end{eqnarray*}
where $1\leq j\leq J,1\leq k\leq K$ and $0\leq r\leq 3$. Estimates of $%
\theta _{jk},\phi$ and $\phi _{j}^{(0)}$ can then be obtained by
substituting the above estimates of $(\varphi,\pi )=(\alpha
_{(0),jk},\beta_{(1),j+},\break\gamma_{(2),+k},\pi_{r})$ into equation
 (\ref{transform}). This yields
\begin{eqnarray}
\label{E:solution1}\qquad\hat{\theta}_{jk}&=&\biggl(\frac{n_{(0),jk}}{n_{(0),j+}}\biggr)
\biggl( \frac{%
n_{(0),j+}+n_{(1),j+}}{n_{0}+n_{1}} \biggr),  \\
 \hat{\phi}&=&1-\hat{\pi}_{0}-\hat{\pi}_{1}, \nonumber \\
\label{E:solution3} \hat{\phi}_{j}^{(0)}&=&\frac{\hat{\pi}_{1}\hat{\beta}_{(1),j+}}
{\hat{\pi}_{0}%
\hat{\alpha}_{(0),j+}+\hat{\pi}_{1}\hat{\beta}_{(1),j+}}.
\end{eqnarray}
Estimates of \{${\phi}_{j}^{(1)},j=1,\ldots,J$\} can be
obtained as solutions of the following $K$ simultaneous equations,
provided they are in the parameter space:
\begin{equation}\label{E:solution4}
\sum_{j=1}^{J}\bigl(1-\hat{\phi}_{j}^{(1)}\bigr)\hat{\theta}_{jk}=
\frac{\hat{\pi}_{2}}{1-\hat{\pi}_{0}-\hat{\pi}_{1}}\hat{\gamma}_{(2),+k}.
\end{equation}
This approach yields ML estimates,
providing the estimates lie within the parameter space,
that is, the probabilities lie between zero and one. The expressions for $\hat{\theta}%
_{jk},\hat{\phi}$ and $\hat{\phi}_{j}^{(0)}$ always yield estimates in $%
[0,1] $. The equations in (\ref{E:solution4}), however, may or may not
yield solutions for $\{\phi _{j}^{(1)}\}$ that lie in $[0,1]$. If they
do, then
estimates from this approach are ML estimates and the ML estimates of $%
\theta _{jk}$, $\phi $ and $\phi _{j}^{(0)}$ are unique. If not, this
approach fails to yield ML estimates of the parameters of interest. In
this case, however, the EM algorithm can still be used, and whether the
ML estimate is unique or not depends on the form of the likelihood. If
the
likelihood is unimodel, the ML estimate is unique. The solution set for (\ref{E:solution4})
depends on whether $J=K$ or $J>K$. When $J=K$ there are
$J$
equations for $J$ unknowns. Provided the $J\times J$ matrix, $\hat{\Theta}=(%
\hat{\theta}_{jk})$, is nonsingular, these equations yield a unique
solution that may or may not lie in the parameter space. When $J\geq K$ and $%
\hat{\Theta}$ has rank $K^{\prime }<J$, the solution set is a linear
subspace of dimension $J-K^{\prime }$. If the solution space intersects
the parameter space $[0,1]^{J}$, then this approach yields the ML
estimates. For example, consider the case where $J=3$, $K=2$ and $\hat{\Theta%
}$ is of full rank $K$, the solution set to (\ref{E:solution4}) is a
straight line. When it intersects the unit cube representing the
parameter
space, this approach yields unique ML estimates of $\theta _{jk},\phi $ and $%
\phi _{j}^{(0)}$, but any point in $[0,1]^{J}$ that is in the solution
set of (\ref{E:solution4}) is a ML estimate for $\{\phi _{j}^{(1)}\}$.
However, when the solution set does not intersect the unit cube, this
method fails to yield the ML estimates of the parameters. The EM
algorithm can be implemented to find ML estimates, which may or may not
be unique. When $J<K$, noniterative ML estimates do not exist and the
EM algorithm can be applied to compute ML estimates.

\begin{table*}[t]
\caption{$2 \times 2$ tables with supplemental margins for both
variables}\label{t3}
\begin{tabular}{@{}cc@{}}
{\begin{tabular}{@{}ccccc@{}}
\multicolumn{1}{@{}l}{3A}&&&&\\
\hline
& &\multicolumn{3}{c}{$\bolds{Z_{(2)}}$}\\
\hhline{~~---}\\[-5pt]
 & & \textbf{1} & \textbf{2} & \textbf{Missing}\\\hline
& \textbf{1} & 50 & 150 & 30 \\
$\bolds{Z_{(1)}}$& \textbf{2} & 75 & \phantom{0}75 & 60 \\
& \textbf{Missing} & 28 & \phantom{0}60 & 50\\\hline
\end{tabular}}&
{\begin{tabular}{@{}ccccc@{}}
\multicolumn{1}{@{}l}{3B}&&&&\\
\hline
& &\multicolumn{3}{c}{$\bolds{Z_{(2)}}$}\\
\hhline{~~---}\\[-5pt]
& & \textbf{1} & \textbf{2} & \textbf{Missing}\\\hline
& \textbf{1} & 100 & 50 & 30 \\
$\bolds{Z_{(1)}}$& \textbf{2} & \phantom{0}75 & 75 & 60 \\
& \textbf{Missing} & \phantom{0}28 & 60 & 50\\\hline
\end{tabular}}
\end{tabular}
\end{table*}

The closed-form estimates (\ref{E:solution1}) of $\theta$ are simply
the product of the estimated conditional probabilities of $Z_{(2)}=k$
given $Z_{(1)}=j$ from the complete
cases and the marginal probabilities of $Z_{(1)}=j$ from the cases with $%
Z_{(1)}$ observed. These estimates maximize the block-monotone reduced
likelihood discussed in Section \ref{2}, which drops the data for $Z_{(2)}$
from the pattern $P_{2}$ with $Z_{(2)}$ observed and $Z_{(1)}$ missing.
One would expect the data in $P_{2}$ to provide additional information
for the marginal distribution of $Z_{(2)}$, but this is only the case
if the data in
$P_{2}$ are inconsistent with the data on $Z_{(2)}$ from $P_{0}$ and $P_{1}$%
, in the sense of yielding estimates of $\{\phi _{j}^{(1)}\}$ from
(\ref{E:solution4}) that lie outside the interval [0, 1].

\section{a Restricted BCMAR Model}\label{4}

In the unrestricted BCMAR model (\ref{E:mechanism}), the missingness of
$Z_{(2)}$ is allowed to depend not only on the (perhaps unobserved)
value of $Z_{(1)}$ but also on whether $Z_{(1)}$
is missing or not. If, given the value of $Z_{(1)}$, the probability of $%
Z_{(2)}$ being missing is assumed the same for the cases with $Z_{(1)}$
observed and missing, we then have the restricted BCMAR model:
\begin{eqnarray}\label{E:mechanism1}
 \hspace*{15pt}P\bigl(M_{(1)}=1|Z_{(1)}=j,Z_{(2)}=k\bigr)&=&\phi, \nonumber \hspace*{-15pt}
 \\[-8pt]\\[-8pt]
 \hspace*{15pt}P\bigl(M_{(2)}=1|M_{(1)}=l,Z_{(1)}=j,Z_{(2)}=k\bigr)&=&\phi _{j},\hspace*{-15pt}
 \nonumber
\end{eqnarray}
where $l=1,2$ and $1\leq j\leq J,1\leq k\leq K$. The number of the
parameters in this model is $JK+J$ which is always less than the degree
of freedom $JK+J+K$ in the data. The explicit estimates in
(\ref{E:solution1}) are no longer ML estimates of $\{\theta _{jk}\}$,
and EM is needed to obtain ML estimates of the parameters. In the E
step, the partially classified observations are effectively distributed
into the table according to the corresponding estimated probabilities:
\begin{eqnarray*}
n_{(1),jk}^{(t)}&=& n_{(1),j+}\cdot \frac{\theta _{jk}^{(t)}}{\theta _{j+}^{(t)}}, \\
n_{(2),jk}^{(t)}&=& n_{(2),+k}\cdot \frac{(1-{\phi _{j}}%
^{(t)})\theta _{jk}^{(t)}}{\sum_{j=1}^{J}(1-{\phi _{j}}^{(t)})\theta
_{jk}^{(t)}}, \\
n_{(3),jk}^{(t)}&=& n_{(3),++}\cdot \frac{{\phi _{j}}^{(t)}\theta
_{jk}^{(t)}}{\sum_{j=1}^{J}{\phi _{j}}^{(t)}\theta _{j+}^{(t)}}.
\end{eqnarray*}
In the M step, new estimates are calculated as
\[
 \theta _{jk}^{(t+1)}= \frac{n_{(0),jk}+n_{(1),jk}^{(t)}+n_{(2),%
jk}^{(t)}+n_{(3),jk}^{(t)}}{n},
\]
\[
\phi =\frac{n_{2}+n_{3}}{n},
\]
\begin{eqnarray*}
&&{\phi _{j}}^{(t+1)}
\\
&&\quad= \frac{\sum_{k}n_{(1),jk}^{(t)}+\sum_{k}n_{(3),%
jk}^{(t)}}{n_{(0),j+}+\sum_{k}n_{(1),jk}^{(t)}+\sum_{k}n_{(2),%
jk}^{(t)}+\sum_{k}n_{(3),jk}^{(t)}}.
\end{eqnarray*}
The E step and M step alternate until the parameter estimates converge.
Since $\phi $ is estimable directly and is
unchanged throughout the EM algorithm, starting values are only needed for $%
\{\theta _{jk}\}$ and $\{\phi _{j}\}$. Complete-case estimates or
pooled estimates from the monotone pattern $P_{0}$ and $P_{1}$ can be
used as starting values of $\{\theta _{jk}\}.$ Estimates of $\{\phi
_{j}^{(0)}\}$ in (\ref{E:solution3}) or any constant in $(0,1)$ can be
taken as initial values of $\{\phi _{j}\}$.

The restricted BCMAR model (\ref{E:mechanism1}) is a submodel of the
unrestricted BCMAR model (\ref{E:mechanism}) obtained by assuming $\phi
_{j}^{(0)}=\phi _{j}^{(1)}$. The restricted model is plausible when the
mechanism of missingness of $Z_{(1)}$ is relatively unrelated to the
mechanism of missingness of $Z_{(2)}$, so the probability that one variable
is missing is not thought to be related to whether the other variable
is missing. The appeal of the restricted model is that it is more
parsimonious and will tend to yield more efficient estimates of the
parameters of interest. A likelihood ratio test can be applied to test
the restricted \mbox{BCMAR} assumption against the more general unrestricted
\mbox{BCMAR} model, and one may favor the restricted \mbox{BCMAR} if this test is not
rejected.

\section{Numerical Examples}\label{5}

\subsection{Examples with $J=K=2$}\label{51}

For data given in the $2\times 2$ Table \ref{t3}A with supplemental margins,
the noniterative estimates of $\{\theta _{jk}\}$ that drop the data in
$P_{2}$ are ML estimates under the unrestricted BCMAR model. The
estimates of $\{\theta _{jk}\}$ are also close to those in the
restricted BCMAR and MAR models which involve all the data
(Table \ref{t4}). However, for data in Table \ref{t3}B, the marginal distribution of $%
Z_{(2)}$ in $P_{2}$ is substantially different from that in the
monotone pattern $P_{0}$ and $P_{1}$. In this case, the unrestricted
\mbox{BCMAR} model yields the estimates of $\{\phi _{j}^{(1)}\}$ from
(\ref{E:solution4}) that do not lie between 0 and 1. The EM algorithm
applied to all the data is needed to obtain the ML estimates, and the
estimates of $\{\theta _{jk}\}$ are different from those in the
restricted \mbox{BCMAR} and MAR models (Table \ref{t5}).

\begin{table*}
\caption{Estimates of parameters for data in Table \protect\ref{t3}A}\label{t4}
\begin{tabular}{@{}lcccc@{}c@{\hspace*{10pt}}ccccc@{}}
\hline
& \multicolumn{4}{c@{}}{\textbf{Parameter of interest}} && \multicolumn{5}{@{}c@{}}
{\textbf{Nuisance parameter}} \\
\hhline{~----~-----}\\[-5pt]
& $\bolds{\theta_{11}}$ & $\bolds{\theta_{12}}$ & $\bolds{\theta_{21}}$ &
$\bolds{\theta_{22}}$ && $\bolds{\phi}$ & $\bolds{\phi_{1}^{(0)}}$ &
$\bolds{\phi_{2}^{(0)}}$ & $\bolds{\phi_{1}^{(1)}}$ & $\bolds{\phi_{2}^{(1)}}$ \\
\hline
{Unrestricted BCMAR} & \multicolumn{4}{c}{} & & && & & \\
noniterative estimate & 0.131 & 0.392 & 0.239 & 0.239 && 0.239 & 0.130 &0.286 & 0.113 & 0.636 \\
EM algorithm & 0.131 & 0.392 & 0.239 & 0.239 && 0.239 & 0.130 & 0.286 &0.113 & 0.636
\\[3pt]
{Restricted BCMAR} & \multicolumn{4}{c}{} & & & \multicolumn{4}{c}{$\phi_j^{(0)}=
\phi_j^{(1)},j=1,2$} \\
 &\multicolumn{4}{c}{} && $\phi$ & \multicolumn{2}{c}{$\phi_1$} &\multicolumn{2}{c}{$\phi_2$} \\
EM algorithm & 0.126 & 0.390 & 0.238 & 0.246 && 0.239 &\multicolumn{2}{c}{ 0.157} &
\multicolumn{2}{c}{0.333} \\[3pt]
{Restricted MAR} & \multicolumn{4}{c}{} && \multicolumn{3}{c}{}&
\multicolumn{2}{c}{$\phi_1^{(1)}=\phi_2^{(1)}$} \\
&\multicolumn{4}{c}{} && $\phi$ & $\phi_1^{(0)}$ & $\phi_2^{(0)}$ &
\multicolumn{2}{c}{$\phi^{(1)}$}\\
EM algorithm & 0.127 & 0.398 & 0.232 & 0.243 && 0.239 & 0.130 & 0.286 &
\multicolumn{2}{c}{0.362} \\
\hline
\end{tabular}
\end{table*}

\begin{table*}[b]
\caption{Estimates of parameters for data in Table
\protect\ref{t3}B}\label{t5}
\begin{tabular}{@{}lcccc@{}c@{\hspace*{10pt}}ccccc@{}}
\hline
& \multicolumn{4}{c@{}}{\textbf{Parameters of interest}} &&
\multicolumn{5}{@{}c@{}}{\textbf{Nuisance parameter}} \\
\hhline{~----~-----}\\[-5pt]
& $\bolds{\theta_{11}}$ & $\bolds{\theta_{12}}$ & $\bolds{\theta_{21}}$ &
$\bolds{\theta_{22}}$ && $\bolds{\phi}$ & $\bolds{\phi_{1}^{(0)}}$ & $\bolds{\phi_{2}^{(0)}}$ &
$\bolds{\phi_{1}^{(1)}}$ & $\bolds{\phi_{2}^{(1)}}$ \\
\hline
{Unrestricted BCMAR} & \multicolumn{4}{c}{} && & & & & \\
noniterative estimate & 0.308 & 0.154 & 0.269 & 0.269 && 0.261 & 0.167 & 0.286 & 2.507 &
$-$1.476 \\
EM algorithm & 0.297 & 0.153 & 0.236 & 0.314 && 0.261 & 0.167 & 0.286 & 0.867 & 0 \\[3pt]
{Restricted BCMAR} & \multicolumn{4}{c}{} & & &
\multicolumn{4}{c@{}}{$\phi_j^{(0)}=\phi_j^{(1)},j=1,2$} \\
&\multicolumn{4}{c}{} && $\phi$ & \multicolumn{2}{c}{$\phi_1$} &
\multicolumn{2}{c@{}}{$\phi_2$} \\
EM algorithm & 0.274 & 0.175 & 0.242 & 0.309 && 0.261 &\multicolumn{2}{c}{ 0.197} &
\multicolumn{2}{c@{}}{0.320} \\[3pt]
{Restricted MAR} & \multicolumn{4}{c}{} & \multicolumn{3}{c}{}&
\multicolumn{2}{c@{}}{$\phi_1^{(1)}=\phi_2^{(1)}$}
\\[1pt]
&\multicolumn{4}{c}{} & &$\phi$ & $\phi_1^{(0)}$ & $\phi_2^{(0)}$ &
\multicolumn{2}{c@{}}{$\phi^{(1)}$} \\
EM algorithm & 0.279 & 0.174 & 0.239 & 0.308 && 0.261 & 0.167 & 0.286 &
\multicolumn{2}{c@{}}{0.362} \\
\hline
\end{tabular}
\end{table*}

\subsection{Examples with $J=3,K=2$}\label{52}

\begin{figure}[b]

\includegraphics{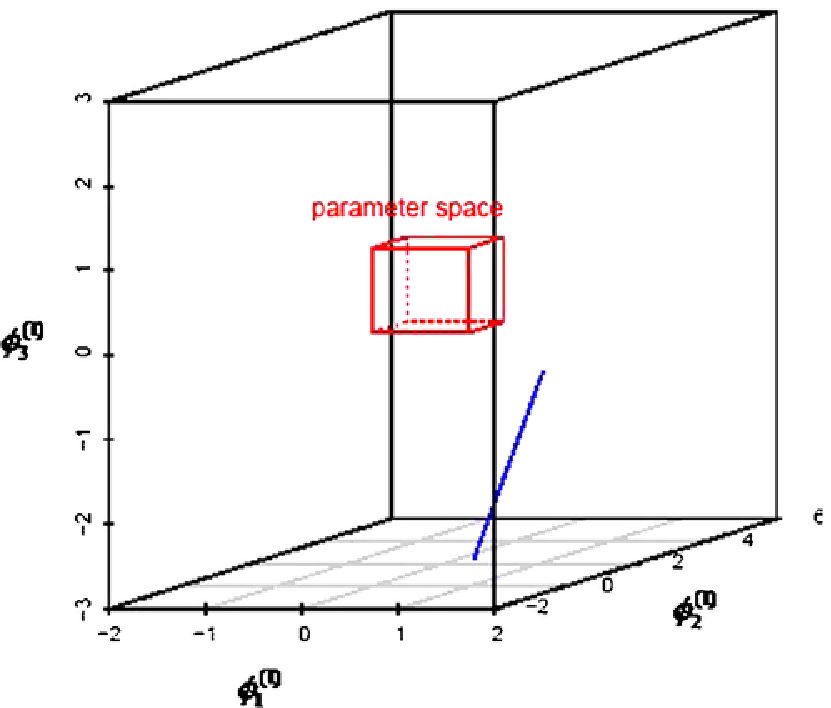}

\caption{Noniterative estimates of $\phi_{j}^{(1)}$ for
data in Table \protect\ref{t6}A.}\label{f1}
\end{figure}

\begin{figure}[b]

\includegraphics{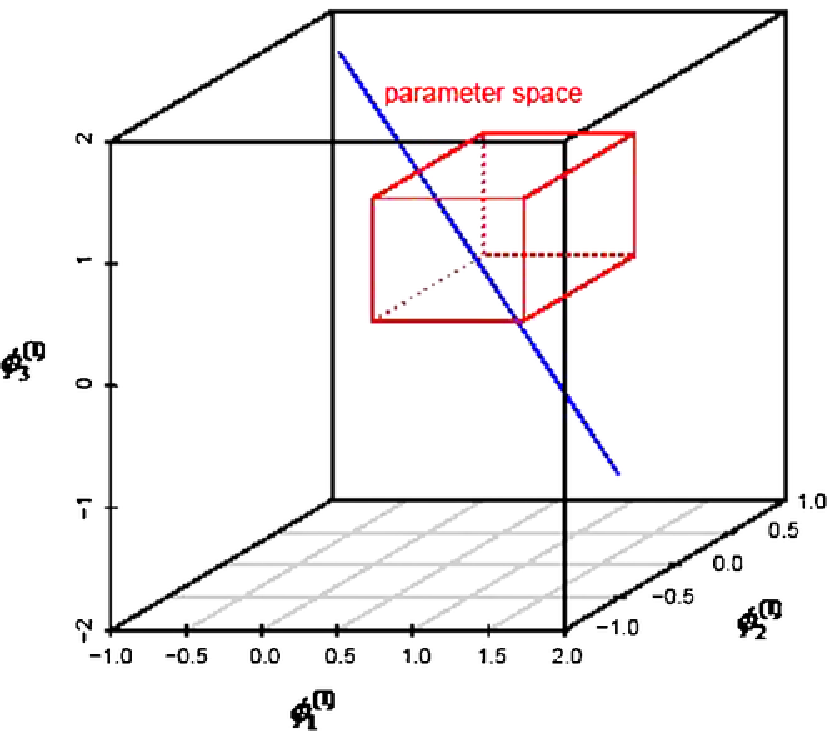}

\caption{Noniterative estimates of $\phi_{j}^{(1)}$ for
data in Table \protect\ref{t6}B.}\label{f2}
\end{figure}

Table \ref{t6}A and B give data for the case $%
J=3$, $K=2$ for which the solution set to (\ref{E:solution4}) is a
straight line. The parameter space for $\{\phi _{j}^{(1)}\}$ is a unit
cube, as displayed in Figures \ref{f1} and \ref{f2}. For the data in Table \ref{t6}A, the
solution line does not intersect the cube (Figure \ref{f1}), so ML estimates
in the unrestricted BCMAR model are obtained iteratively using all the
data (Table \ref{t7}). For the data in Table \ref{t6}B, the marginal distribution of
$Z_{(2)}$ in $P_{2}$ is similar to that in $P_{0}$ and $P_{1}$ and the
solution line intersects the cube (Figure \ref{f2}), and the noniterative
estimates obtained by dropping the data in $P_{2}$, displayed in Table
\ref{t8}, are the ML estimates of $\{\theta
_{jk}\}$, although there are multiple ML estimates for $\{\phi _{j}^{(1)}\}$%
. ML estimates in the restricted BCMAR and MAR models are unique for
both data sets in Table \ref{t6}.

\begin{table*}
\caption{$3 \times 2$ tables with supplemental margins for both
variables}\label{t6}
\begin{tabular}{@{}cc@{}}
{\begin{tabular}{@{}ccccc@{}}
\multicolumn{1}{@{}l}{6A}&&&&\\
\hline
& &\multicolumn{3}{c}{$\bolds{Z_{(2)}}$}\\
\hhline{~~---}\\[-5pt]
 & & \textbf{1} & \textbf{2} & \textbf{Missing}\\\hline
&\textbf{ 1} & 100 & 50 & 30 \\
$\bolds{Z_{(1)}}$& \textbf{2} & \phantom{0}75 & 75 & 60 \\
& \textbf{3} & \phantom{0}32 & 67 & 20 \\
& \textbf{Missing} & \phantom{0}28 & 60 & 50\\\hline
\end{tabular}}&
{\begin{tabular}{@{}ccccc@{}}
\multicolumn{1}{@{}l}{6B}&&&&\\
\hline
& &\multicolumn{3}{c}{$\bolds{Z_{(2)}}$}\\
\hhline{~~---}\\[-5pt]
& & \textbf{1} & \textbf{2} & \textbf{Missing}\\\hline
& \textbf{1} & 50 & 150 & 30 \\
$\bolds{Z_{(1)}}$& \textbf{2} & 75 & \phantom{0}75 & 60 \\
& \textbf{3} & 32 & \phantom{0}67 & 20 \\
& \textbf{Missing} & 28 & \phantom{0}60 & 50\\\hline
\end{tabular}}
\end{tabular}
\end{table*}

\section{Muscatine Coronary Risk Factor Study}\label{6}

The Muscatine Coronary Risk Factor
Study (MCRF) is a longitudinal study of obesity in 4856 school
children. Five cohorts (ages 5--7, 7--9, 9--11, 11--13, 13--15) of boys and
girls were measured for height and weight in 1977, 1979 and 1981.
Children with relative weight greater than 110 percent of the median
weight for their age-gender-height group were classified as obese, and
at any time point about 20 percent of the children were obese. We are
interested in estimating obesity rates over time and evaluating whether
or not these rates differ by gender. The study was first presented by
Woolson and Clarke (\citeyear{r36}), and further analyses can be found
in, for example, Baker (\citeyear{r1}), Ekholm and Skinner
(\citeyear{r8}), Lipsitz, Parzen and Molenberghs (\citeyear{r16}) and
Birmingham and Fitzmaurice (\citeyear{r3}).

\begin{table*}[t]
\tabcolsep=0pt
\caption{Estimates of parameters for data in Table
\protect\ref{t6}A}\label{t7}
\begin{tabular*}{\textwidth}{@{\extracolsep{4in minus 4in}}lcccccccccccccc@{}}
\hline
& \multicolumn{6}{c@{}}{\textbf{Parameter of interest}} &&
\multicolumn{7}{@{}c@{}}{\textbf{Nuisance parameter}} \\
\ccline{2-7,9-15}\\[-5pt]
& $\bolds{\theta_{11}}$ & $\bolds{\theta_{12}}$ & $\bolds{\theta_{21}}$ &
$\bolds{\theta_{22}}$ & $\bolds{\theta_{31}}$ & $\bolds{\theta_{32}}$ &&
$\bolds{\phi}$ & $\bolds{\phi_{1}^{(0)}}$ &
$\bolds{\phi_{2}^{(0)}}$ & $\bolds{\phi_{3}^{(0)}}$ & $\bolds{\phi_{1}^{(1)}}$ &
$\bolds{\phi_{2}^{(1)}}$ & \multicolumn{1}{c@{}}{$\bolds{\phi_{3}^{(1)}}$}\\
\hline
{Unrestricted BCMAR}\\
Noniterative estimate & 0.236 & 0.118 &0.206 & 0.206 & 0.076 & 0.158 && 0.213 & 0.167 &
0.286 & 0.168 &
\multicolumn{3}{c@{}}{\textit{no solution in $[0,1]^{3}$}} \\
EM algorithm & 0.235 & 0.117 & 0.192 & 0.219 & 0.071 & 0.166 && 0.213& 0.167 & 0.286 &
0.168 & 1 & \hspace*{5pt}0.037 &
\multicolumn{1}{c@{}}{0}\\[3pt]
{Restricted BCMAR}&\multicolumn{6}{c}{}&&
\multicolumn{7}{c}{$\phi_j^{(0)}=\phi_j^{(1)},j=1,2,3$}\\
&\multicolumn{6}{c}{}&&$\phi$&\multicolumn{2}{c}{$\phi_1$}&
\multicolumn{2}{c}{$\phi_2$}&\multicolumn{2}{c@{}}{$\phi_3$}\\
EM algorithm & 0.218 & 0.126 & 0.194 & 0.224 & 0.069 & 0.168 &&0.213&
\multicolumn{2}{@{}c}{0.196}&\multicolumn{2}{c}{0.322}&\multicolumn{2}{c@{}}{0.190} \\[3pt]
{Restricted MAR}&\multicolumn{6}{c}{}&
&\multicolumn{4}{c}{}&\multicolumn{3}{c@{}}{$\phi_1^{(1)}=\phi_2^{(1)}=\phi_3^{(1)}$}\\
&\multicolumn{6}{c}{}&&$\phi$&$\phi_1^{(0)}$&$\phi_2^{(0)}$&$\phi_3^{(0)}$&
\multicolumn{3}{c@{}}{$\phi^{(1)}$}\\
EM algorithm & 0.221 & 0.127 & 0.190 & 0.223&0.070&0.169&&0.213&0.167&0.286&0.168&
\multicolumn{3}{c@{}}{0.362}\\
\hline
\end{tabular*}\vspace*{-3pt}
\end{table*}

The analysis is complicated by the study design. Both cross-sectional
and longitudinal information about age trends in obesity rates were
present in~the data. Due to cohort effects, cross-sectional age trends
in obesity rates may be different from longitudinal trends. Ekholm and
Skinner (\citeyear{r8}) found no \mbox{statistical} evidence of cohort
effects. Therefore, in our analyses, cohort effects are assumed
negligible and data are pooled across five age-group cohorts. In~order~to
simplify the illustration, we only use the data from the surveys of
years 1977 and 1981 (Table~\ref{t9}).

The analysis is further complicated by the substantial nonresponse.
Only 40 percent of children provided complete records in 1977 and 1981.
In addition to the complete records, there are three nonresponse
patterns, specifically, two patterns with one missing response and one
pattern with two missing responses. Baker (\citeyear{r1}) reported two
main reasons for nonresponse: (1)~no parental consent form was received
and (2)~the child was not in school on the examination day. For girls,
the missingness of obese status in 1981 is found to depend on the
missingness in 1977 using a chi-square test ($p$-value $<$ 0.0001).
Furthermore, girls measured and classified as obese in 1977 were more
likely to have missing data in 1981 than those classified as nonobese
($p$-value $<$ 0.0001 based on a chi-square test). The estimates of
girls' obesity rates and missing probabilities in the BCMAR model
discussed above are presented in Table \ref{t10}.
For the unrestricted BCMAR model, the estimate from (\ref{E:solution4}) of $%
\{\phi _{1}^{(1)},\phi _{2}^{(1)}\}$ is $(0.274,0.121)$, which is in
the parameter space, so closed form estimates of the parameters are
available. A bootstrap approach was used to estimate standard errors.
If a bootstrap
sample leads to the solutions of $\{\phi _{j}^{(1)}\}$ from (\ref%
{E:solution4}) that lie outside of the parameter space, the EM
algorithm is used to obtain the ML estimates. Among the 1000 bootstrap
samples, 23.2\%
of the samples yield the solutions of $\{\phi _{j}^{(1)}\}$ from (\ref{E:solution4}%
) that are outside of the parameter space.

\begin{table*}[t]
\tabcolsep=0pt
\caption{Estimates of parameters for data in Table
\protect\ref{t6}B}\label{t8}
\begin{tabular*}{501pt}{@{\extracolsep{\fill}}lcccccccccccccc@{}}
\hline & \multicolumn{6}{c@{}}{\textbf{Parameter of interest}} &&
\multicolumn{7}{@{}c@{}}{\textbf{Nuisance parameter}} \\
\ccline{2-7,9-15}\\[-5pt]
& $\bolds{\theta_{11}}$ & $\bolds{\theta_{12}}$ & $\bolds{\theta_{21}}$ &
$\bolds{\theta_{22}}$ & $\bolds{\theta_{31}}$ & $\bolds{\theta_{32}}$ &&
$\bolds{\phi}$ & $\bolds{\phi_{1}^{(0)}}$ &
$\bolds{\phi_{2}^{(0)}}$ & $\bolds{\phi_{3}^{(0)}}$ & $\bolds{\phi_{1}^{(1)}}$ &
$\bolds{\phi_{2}^{(1)}}$ & $\bolds{\phi_{3}^{(1)}}$\\
\hline
{Unrestricted BCMAR} & \multicolumn{6}{c}{}\\
Noniterative estimate & 0.103 & 0.309 & 0.188 & 0.188 & 0.069 & 0.144 && 0.198 & 0.130 &
0.286 & 0.168 & \multicolumn{3}{c@{\hspace*{-10pt}}}{\textit{multiple solutions in $[0,1]^{3}$}} \\
EM algorithm & 0.103 & 0.309 & 0.188 & 0.188 & 0.069 & 0.144 && 0.198 & 0.130 & 0.286 &
0.168 & \multicolumn{3}{c@{}}{\textit{multiple solutions}} \\[3pt]
{Restricted BCMAR}&\multicolumn{6}{c}{}&&
\multicolumn{7}{c}{$\phi_j^{(0)}=\phi_j^{(1)},j=1,2,3$}\\
&\multicolumn{6}{c}{}&&$\phi$&\multicolumn{2}{c}{$\phi_1$}&\multicolumn{2}{c}{$\phi_2$}&
\multicolumn{2}{c@{}}{$\phi_3$}\\
EM algorithm & 0.100 & 0.307 & 0.189 & 0.193 & 0.067 & 0.144 && 0.198&
\multicolumn{2}{c}{0.154}&\multicolumn{2}{c}{0.328}&\multicolumn{2}{c@{}}{0.197} \\[3pt]
{Restricted MAR}&\multicolumn{6}{c}{}&&\multicolumn{4}{c}{}&
\multicolumn{3}{c@{}}{$\phi_1^{(1)}=\phi_2^{(1)}=\phi_3^{(1)}$}\\
&\multicolumn{6}{c}{}&&$\phi$&$\phi_1^{(0)}$&$\phi_2^{(0)}$&$\phi_3^{(0)}$&
\multicolumn{3}{c@{}}{$\phi^{(1)}$}\\
EM algorithm & 0.101 & 0.311 & 0.184 & 0.190&0.068&0.146&&0.198&0.130&0.286&0.168&
\multicolumn{3}{c@{}}{0.362}\\
\hline
\end{tabular*}
\end{table*}

Likelihood ratio tests can be utilized to test the two submodels
discussed above against the more general unrestricted BCMAR model.
Denote the unrestricted BCMAR model as M1, the restricted BCMAR
model as M2 and the restricted MAR model in Section~\ref{3} as M3, and let $l_{%
\max}$ represent the maximized value of the loglikelihood. We find that\break
$-2(l_{\max}(M2)-l_{\max}(M1))=-2(-4569.823+\break 4535.292)=69.062 $, which
yields a $p$-value $<0.0001$ when compared to $\chi _{2}^{2}$. There is
strong evidence that the
restricted \mbox{BCMAR} model does not fit the data. On the other hand, $l_{%
\max}(M3)$ is close to $l_{\max}(M1)$, and we cannot differentiate the
restricted MAR model from the unrestricted \mbox{BCMAR}
model.

\begin{table}[t]
\tablewidth=154pt
\caption{Tables of data from muscatine coronary risk factor
study}\label{t9}
\begin{tabular}{@{}lcccc@{}}
\hline
& &\multicolumn{3}{c@{}}{\textbf{1981}}\\
\hhline{~~---}\\[-5pt]
& & \textbf{1} & \textbf{2} & \textbf{Missing}\\
\hline
\textbf{Girls}&&&&\\
  & \textbf{1} & 701 & \phantom{0}98 & 497 \\
\textbf{1977}& \textbf{2} & \phantom{0}59 & 111 & 183 \\
& \textbf{Missing} & 408 & 139 & 174\\
\hline
\textbf{Boys} & &&&\\
  & \textbf{1} & 699 & \phantom{0}98 & 566 \\
\textbf{1977}& \textbf{2} & \phantom{0}72 & 116 & 141 \\
& \textbf{Missing} & 473 & 125 & 196\\\hline
\end{tabular}
\tabnotetext[]{}{\textit{Notes}: 1${}={}$not obese, 2${}={}$obese.}
\vspace*{-2pt}
\end{table}

Similarly for the boys, the estimate from (%
\ref{E:solution4}) of $\{\phi _{1}^{(1)},\phi _{2}^{(1)}\}$ in the
unrestricted BCMAR model is $(0.228,0.325)$, which is in the parameter
space, and closed form estimates of the parameters are available. Among
1000 bootstrap samples, only 28 samples yield the solutions of $\{\phi
_{j}^{(1)}\}$ from (\ref{E:solution4}) outside of the parameter space.
The likelihood ratio test yields strong evidence against the restricted
BCMAR model, with
$-2(l_{\max}(M2)-l_{\max}(M1))=\break-2(-4748.48+4713.03)=70.9$ on two
degrees of freedom. On the other hand, $l_{\max}(M3)$ is close to
$l_{\max}(M1)$, and the restricted MAR model seems to be satisfactory
(Table~\ref{t11}).

\begin{table*}[t]
\caption{Estimates of girls' obesity rates}\label{t10}
\tabcolsep=3pt
\begin{tabular*}{\textwidth}{@{\extracolsep{\fill}}lcccccccccc@{}}
\hline
& \multicolumn{4}{c}{\textbf{Obesity rate}} &\multicolumn{5}{c}{\textbf{Nuisance parameter}} &
\multirow{2}{60pt}{\textbf{Observed data}} \\
\hhline{~~--~~---~~}\\[-5pt]
& $\bolds{\theta_{11}}$ & $\bolds{\theta_{12}}$ & $\bolds{\theta_{21}}$ & $\bolds{\theta_{22}}$
&$\bolds{\phi}$ & $\bolds{\phi_{1}^{(0)}}$ & $\bolds{\phi_{2}^{(0)}}$ &
$\bolds{\phi_{1}^{(1)}}$& $\bolds{\phi_{2}^{(1)}}$&\textbf{loglikelihood}\\
\hline
{Complete-case estimate} & 0.723 & 0.101 & 0.061 & 0.115 &--&--&--&--&--&--\\
& (0.014) & (0.010) & (0.008) & (0.010) & & & & & &\\[3pt]
{Restricted MAR} & \multicolumn{7}{c}{}&\multicolumn{2}{c}{$\phi_1^{(1)}=\phi_2^{(1)}$}&\\
& & & & & $\phi$ & $\phi_{1}^{(0)}$ & $\phi_{2}^{(0)}$ & \multicolumn{2}{c}{$\phi^{(1)}$}&\\
EM algorithm & 0.685 & 0.099 & 0.073 & 0.143 & 0.304 & 0.383 & 0.518 &
\multicolumn{2}{c}{0.241}&\\
& (0.012) & (0.009) & (0.009) & (0.010) & (0.010)& (0.006) & (0.023)&
\multicolumn{2}{c}{(0.016)} &
$-$4535.605\\[3pt]
{Restricted BCMAR} &\multicolumn{5}{c}{}&\multicolumn{4}{c}{$\phi_j^{(0)}=\phi_j^{(1)}, j=1,2$}&\\
& & & & & $\phi$ & \multicolumn{2}{c}{$\phi_{1}$} & \multicolumn{2}{c}{$\phi_{2}$}&\\
EM algorithm & 0.683 & 0.103 & 0.070 & 0.143 & 0.304 & \multicolumn{2}{c}{0.335}&
\multicolumn{2}{c}{0.455}&\\
& (0.011) & (0.009) & (0.008) & (0.010) & (0.010)&\multicolumn{2}{c}{(0.006)} &
\multicolumn{2}{c}{(0.022)}& $-$4569.823\\[3pt]
{Unrestricted BCMAR} & & & & & $\phi$ & $\phi_{1}^{(0)}$ & $\phi_{2}^{(0)}$&
$\phi_{1}^{(1)}$&$\phi_{2}^{(1)}$&\\
noniterative estimate & 0.690 & 0.096 & 0.074 & 0.140 & 0.304 & 0.383 & 0.518 & 0.274 & 0.121&\\
& (0.012) & (0.010) & (0.009) & (0.010) & (0.010)& (0.006) & (0.023) &
(0.034) & (0.122) & $-$4535.292\\
\hline
\end{tabular*}
\end{table*}

The models considered above show a small effect on the fitted values of
obesity rates and their standard errors. For boys, the marginal
distributions of 1981 obesity rates are quite similar for those with
1977 obesity rates observed or not. If we consider only the cases with
1977 obesity rates observed, the noniterative block-monotone reduced ML
estimates of obesity rates for the unrestricted BCMAR model are ML
estimates, and these are close to ML estimates in the restricted BCMAR
and MAR models. Furthermore, $\hat{\phi}_{1}^{(0)}$ and
$\hat{\phi}_{2}^{(0)}$ are close to one another, which suggests a MCAR
mechanism. As a consequence, complete-case estimates of obesity rates
are also similar to those in three models considered above. For girls,
for the same reason, noniterative block-monotone reduced ML estimates
of obesity rates for the unrestricted BCMAR model are ML estimates and
are close to those in the restricted BCMAR and MAR models. However,
$\hat{\phi}_{1}^{(0)}$ and $\hat{\phi}_{2}^{(0)}$ are quite different,
and, as a consequence, complete-case estimates of obesity rates are not
similar to those in the other three models.

\begin{table*}[b]
\caption{Estimates of boys' obesity rates}\label{t11}
\tabcolsep=3pt
\begin{tabular*}{\textwidth}{@{\extracolsep{\fill}}lcccccccccc@{}}
\hline
& \multicolumn{4}{c}{\textbf{Obesity rate}} &\multicolumn{5}{c}{\textbf{Nuisance parameter}} &
\multirow{2}{60pt}{\textbf{Observed data}} \\
\hhline{~~--~~---~~}\\[-5pt]
& $\bolds{\theta_{11}}$ & $\bolds{\theta_{12}}$ & $\bolds{\theta_{21}}$ & $\bolds{\theta_{22}}$ &
$\bolds{\phi}$ & $\bolds{\phi_{1}^{(0)}}$ & $\bolds{\phi_{2}^{(0)}}$ & $\bolds{\phi_{1}^{(1)}}$&
$\bolds{\phi_{2}^{(1)}}$&\textbf{loglikelihood}\\
\hline
{Complete-case estimate} & 0.710 & 0.099 & 0.073 & 0.118 &--&--&--&--&--&--\\
& (0.015) & (0.010) & (0.008) & (0.010) & & & & & &\\[3pt]
{Restricted MAR} & \multicolumn{7}{c}{}&\multicolumn{2}{c}{$\phi_1^{(1)}=\phi_2^{(1)}$}&\\
& & & & & $\phi$ & $\phi_{1}^{(0)}$ & $\phi_{2}^{(0)}$ & \multicolumn{2}{c}{$\phi^{(1)}$}&\\
EM algorithm & 0.709 & 0.097 & 0.075 & 0.118 & 0.319 & 0.415 & 0.429 & \multicolumn{2}{c}{0.247}&\\
& (0.011) & (0.009) & (0.008) & (0.008) & (0.009)& (0.006) & (0.025)
& \multicolumn{2}{c}{(0.015)} & $-$4713.142\\[3pt]
{Restricted BCMAR} &
\multicolumn{5}{c}{}&\multicolumn{4}{c}
{$\phi_j^{(0)}=\phi_j^{(1)}, j=1,2$}&\\
& & & & & $\phi$ & \multicolumn{2}{c}{$\phi_{1}$} & \multicolumn{2}{c}{$\phi_{2}$}&\\
EM algorithm & 0.709 & 0.098 & 0.075 & 0.118 & 0.319 & \multicolumn{2}{c}{0.360}&
\multicolumn{2}{c}{0.375}&\\
& (0.011) & (0.009) & (0.008) & (0.008) & (0.009)&
\multicolumn{2}{c}{(0.005)} & \multicolumn{2}{c}{(0.023)}& $-$4748.480\\[3pt]
{Unrestricted BCMAR} & & & & & $\phi$ & $\phi_{1}^{(0)}$ & $\phi_{2}^{(0)}$&$\phi_{1}^{(1)}$&
$\phi_{2}^{(1)}$&\\
noniterative estimate & 0.707 & 0.099 & 0.074 & 0.120 & 0.319 & 0.415 & 0.429 & 0.228 & 0.325&\\
& (0.013) & (0.009) & (0.008) & (0.009) & (0.009)& (0.006) & (0.025) &
(0.037) & (0.153) & $-$4713.027\\
\hline
\end{tabular*}
\vspace*{-5pt}
\end{table*}

\section{Two Block bcmar Data with Outcomes from the
Exponential Family Distribution}\label{7}

Suppose, as before, that ${Z}_{(1)}$ takes values $%
1,\ldots,J$ with probabilities $\theta _{j}^{(1)}$ where $\sum {%
\theta _{j}^{(1)}}=1$. The model in Section \ref{3} is generalized here to allow ${%
Z}_{(2)}$ to have an exponential family distribution of full rank.
Thus, we suppose that the density of ${Z}_{(2)}$ given ${Z}_{(1)}$ is
\begin{eqnarray*}
&&f\bigl({Z}_{(2)}|{Z}_{(1)}=j,\theta ^{(2)}\bigr)
\\
&&\quad=a\bigl({Z}_{(2)}\bigr)\exp \bigl[c\bigl({\theta _{j}^{(2)}%
}\bigr)+t\bigl({Z}_{(2)}\bigr)^{T}{\theta _{j}^{(2)}}\bigr],
\end{eqnarray*}%
where $j=1,\ldots,J$, $\theta _{j}^{(2)}$ and ${t(Z}_{(2)})$ are
vectors of dimension $V$, and $c$ is a real-valued function. This
family includes the exponential and normal distribution (with variance known or unknown) as
well as the multivariate normal, normal linear
regression and generalized linear models with canonical links. The mean of $%
t({Z}_{(2)})$ given ${Z}_{(1)}=j$ is given by the $V$-dimensional
vector
\[
\psi _{j}=\psi \bigl(\theta _{j}^{(2)}\bigr)=\frac{\partial c({\theta _{j}^{(2)}})}{%
\partial {\theta _{j}^{(2)}}}.
\]
In a random sample $({Z}_{i(1)},{Z}%
_{i(2)}),i=1,\ldots,n$, the~ML estimate of $\psi _{j}$ is
$\hat{\psi}_{j}=\sum {%
t({Z}_{i(2)})I({Z}_{i(1)}=j)/n_{j+}}$\break where $n_{j+}$ is the number of
observations with ${Z}_{(1)}=j$; the ML estimate of ${\theta
_{j}^{(1)}}$ is
$\hat{\theta}_{j}^{(1)}=n_{j+}/\sum n_{l+}$. The ML estimates of $%
\theta _{j}^{(2)}$ can be obtained from those for~$\psi
_{j}$.

We consider as before the missing data
structure illustrated in Table \ref{t1} with missingness patterns $P_{r}$ with $%
n_{r}$ observations, for $r=0,\ldots,3$. The missingness parameters $%
\Phi=(\phi, \phi_{j}^{(0)},\phi _{j}^{(1)},j=1,\ldots,J)$ are defined
as before in (\ref{E:mechanism}). The parameters in the model are
denoted by the triple~$(\theta ^{(1)},\theta ^{(2)},\Phi )$.

In this case, the likelihood contributions in each~cell from the
(incomplete) data are as follows:
\begin{itemize}
\item For $i\in P_{0}$, the observed data are
${Z}_{i(1)},{Z}_{i(2)},\break
{M}_{i(1)}={M}_{i(2)}=0$ and the likelihood contribution is
proportional to
\begin{eqnarray*}
&&A_{0}\bigl({Z}_{i(1)}=j,{Z}_{i(2)};{\theta ^{(1)},\theta ^{(2)}},\Phi
\bigr) \\
&&\quad={\theta
_{j}^{(1)}}\exp\bigl[c\bigl({\theta _{j}^{(2)}}\bigr)+t\bigl({Z}_{i(2)}\bigr)^{T}
{\theta _{j}^{(2)}}%
\bigr](1-\phi )
\\
&&\qquad{}\cdot\bigl(1-\phi _{j}^{(0)}\bigr).
\end{eqnarray*}
\item For $i\in P_{1}$, the observed data are ${Z}_{i(1)},{M}%
_{i(1)}=0,{M}_{i(2)}=1$, and the likelihood contribution is
proportional to
\begin{eqnarray*}
&&A_{1}\bigl({Z}_{i(1)}=j;{\theta ^{(1)},\theta ^{(2)}},\Phi \bigr)
\\
&&\quad={\theta _{j}^{(1)}}%
(1-\phi )\phi _{j}^{(0)}.
\end{eqnarray*}
\item For $i\in P_{2}$, the observed data are ${Z}_{i(2)},{M}%
_{i(1)}=1,{M}_{i(2)}=0$ and the likelihood contribution is proportional
to
\begin{eqnarray*}
&&A_{2}\bigl({Z}_{(2)};{\theta ^{(1)},\theta ^{(2)}},\Phi \bigr)
\\
&&\quad=\phi \sum_{j=1}^{J}{%
\theta _{j}^{(1)}}\exp \bigl[c\bigl({\theta
_{j}^{(2)}}\bigr)+t\bigl({Z}_{(2)}\bigr)^{T}{\theta
_{j}^{(2)}}\bigr]
\\
&&\qquad\hphantom{\phi \sum_{j=1}^{J}}{}\cdot\bigl(1-\phi _{j}^{(1)}\bigr).
\end{eqnarray*}
\item For $i\in P_{3}$, no elements of ${Z}_{(1)}$ or ${Z}_{(2)}$ are
observed and the data comprise $M_{i(1)}=1,M_{i(2)}=1$. The likelihood
contribution is proportional to
\[
A_{3}\bigl({\theta ^{(1)},\theta ^{(2)}},\Phi \bigr)=\phi
\sum_{j=1}^{J}{\theta _{j}^{(1)}}\phi _{j}^{(1)}.
\]
\end{itemize}

The full observed-data likelihood is then the\break product of such terms and
can be written as $L=L_{0}L_{1}L_{2}L_{3}$, where
\begin{eqnarray*}
L_{0}
&=&
(1-\phi )^{n_{0}}\prod_{j=1}^{J}\bigl\{ \bigl({\theta} _{j}^{(1)}\bigr)^{n_{(0),j+}}
\bigl(1-\phi _{j}^{(0)}\bigr)^{n_{(0),j+}}
\\
&&\hspace{78pt}\cdot
\exp \bigl[c\bigl({\theta _{j}^{(2)}}\bigr)+T_{0j}^{T}{\theta _{j}^{(2)}}\bigr]\bigr\},
\\
L_{1}
&=&
(1-\phi )^{n_{1}}\prod_{j=1}^{J}\bigl\{ \bigl({\theta} _{j}^{(1)}\bigr)^{n_{(1),j+}}
\bigl(\phi _{j}^{(0)}\bigr)^{n_{(1),j+}}\bigr\}, \\
L_{2}
&=&
\phi ^{n_{2}}\prod_{i\in P_{2}}\Biggl\{ \sum_{j=1}^{J}{\theta_{j}^{(1)}}
\bigl(1-\phi _{j}^{(1)}\bigr)
\\
&&\hphantom{\phi ^{n_{2}}\prod_{i\in P_{2}}\Biggl\{ \sum_{j=1}^{J}}{}\cdot
\exp \bigl[c\bigl({\theta _{j}^{(2)}}\bigr)+t\bigl({Z}_{i(2)}\bigr)^{T}{\theta _{j}^{(2)}}\bigr]
\Biggr\},
\\
L_{3}
&=&
\phi ^{n_{3}}\Biggl\{ \sum_{j=1}^{J}{\theta_{j}^{(1)}}\phi_{j}^{(1)}\Biggr\} ^{n_{3}},
\end{eqnarray*}
and $T_{0j}=\sum_{i\in P_{0}}t({Z}_{i(2)})I({Z}_{i(1)}=j)$.

An EM algorithm can readily be applied to maximize the observed-data likelihood. At the E step,
the underlying complete data in patterns $P_{2}$ and $P_{3}$ can be
replaced with their conditional expectations, whereas blocks $P_{0}$
and $P_{1}$ can be treated as complete data. Alternatively, all four
patterns can be incorporated into
the EM approach, with the complete data viewed as all the observations ${Z}%
_{i(1)},{Z}_{i(2)},i=1,\ldots,n$. For the data in block $i\in P_{2}$,
for example, the expectation step involves calculating
\begin{eqnarray*}
&&\!\!\!E\bigl[I\bigl({Z}_{i(1)}=j\bigr)|{Z}_{i(2)},{M}_{i(1)}=1,{M}_{i(2)}=0\bigr]\\
&&\!\!\!\quad=\frac{{\theta_{j}^{(1)}}(1-\phi _{j}^{(1)})\exp [c({\theta _{j}^{(2)}})+
t({Z}_{i(2)})^{T}{%
\theta _{j}^{(2)}}]}{\sum_{l=1}^{J}{\theta _{l}^{(1)}}(1-\phi
_{l}^{(1)})\exp [c({\theta
_{l}^{(2)}})+{t(Z}_{i(2)})^{T}{\theta_{l}^{(2)}}]}.
\end{eqnarray*}
After missing data in each pattern are filled in from the
E step, the M step computes the simple estimates given above for
complete data.

As in the multinomial case, the block-monotone reduced ML estimates of
the parameters $\theta _{j}^{(1)},{ \theta _{j}^{(2)}},\break j=1,\ldots,J$,
are computed from patterns $P_{0},P_{1},$ dropping the data from the
other patterns. The corresponding block-monotone reduced likelihood of
$\theta ^{(1)},\break\theta ^{(2)}$ is
\[
L_{\mathrm{bm}}(P_{0},P_{1})\propto L_{0}\times L_{1},
\]
where the factors in the parameters $\Phi$ can be ignored in
$L_{0},L_{1}$. Unlike the multinomial case, these block-monotone
reduced ML estimates are typically not full ML estimates, since there
is information about the parameters $\theta _{j}^{(2)}$ in the excluded
patterns.

\section{Discussion}\label{8}

Most of the work on MNAR mechanisms concerns selection or
pattern-mixture models, and extensions to include latent random effects
that are applicable to repeated-measures data (Little, \citeyear{r21}).
In this article we consider block-sequential missing data models, where
the variables in the data set are divided into subsets, and the joint
distribution of these variables and their missing data indicators are
factored as a sequence. A characteristic of this class is that
distributions of variables and their missing data indicators are
interleaved, and combinations of selection and pattern-mixture models
can be developed within each block. Except for the work of Robins
(\citeyear{r30}), there appears to be very little existing literature
on missing data mechanisms of this type.

Here we consider a class of block-sequential missing data which we call
block-conditional MAR models, in which missingness in successive blocks
is allowed to depend on observed variables in the block and both
observed and unobserved data in earlier blocks. The proposed class is
related to the models with 2 blocks described in Little and Zhang
(\citeyear{r24}), in the context of regression with missing data. A
block-monotone reduced likelihood approach to estimating these models
is described that yields consistent asymptotically normal estimates
without specifying the distribution of the missing-data mechanism. We
examined here the BCMAR model in some detail for the case of bivariate
categorical data, and showed that maximization of the block-monotone
reduced likelihood can yield fully efficient ML estimates, when
associated estimates of parameters of the missing-data mechanism lie
inside the parameter space. We also discussed more briefly the case
where the variable in the second block comes from an exponential
family, and inference based on the block-monotone reduced likelihood
approach is not in general fully efficient. In future work we plan to
study other BCMAR models involving more than two blocks, continuous and
categorical variables and missing data within each block, and fully
observed covariate information.

The BCMAR model discussed here is related to the ``latent ignorable''
missing data mechanisms proposed to model missing data in the presence
of noncompliance with a treatment (Frangakis and Rubin, \citeyear{r10};
Peng, Little and Raghunathan, \citeyear{r28}). In these cases, there is
a binary compliance variable that indicates whether an individual would
comply with a treatment if assigned to it. In a clinical trial, this
indicator is fully observed for individuals in the active treatment
group, but is completely missing for individuals in the control group,
since they do not have access to the active treatment. The latent
ignorable model assumes MAR within subpopulations defined by the
compliance indicator. Our BCMAR model, applied to that setting,
generalizes this structure by allowing missing data for the stratifying
variable.

The BCMAR model (\ref{BCMAR}) is just one of many possible
block-sequential missing-data models, obtained by placing restrictions
on the parameters of the distributions in each block. Future work might
consider properties of models obtained by imposing other parameter
restrictions, based on plausible assumptions about the nature of the
missing data.

\section*{Acknowledgments}

We appreciate the constructive comments of two referees and an
associate editor which greatly improved the paper.

\end{document}